\documentclass{article}
\usepackage[latin9]{inputenc}
\usepackage{prettyref}
\usepackage{float}
\usepackage{mathtools}
\usepackage{amsmath}
\usepackage{amsthm}
\usepackage{amssymb}
\usepackage{graphicx}
\usepackage{microtype}
\usepackage[unicode=true,
 bookmarks=false,
 breaklinks=false,pdfborder={0 0 1},backref=section,colorlinks=false]
 {hyperref}

\makeatletter

\usepackage{subfigure}

\usepackage[accepted]{icml2025}

\usepackage{amsthm}

\usepackage[capitalize,noabbrev]{cleveref}

\theoremstyle{plain}

\theoremstyle{definition}
\theoremstyle{remark}

\usepackage[textsize=tiny]{todonotes}

\AtBeginDocument{%
\newrefformat{fig}{\hyperref[#1]{Fig.~\ref{#1}}}
}
\AtBeginDocument{%
\newrefformat{app}{\hyperref[#1]{App.~\ref{#1}}}
}

\makeatother

\begin{document}
\global\long\def\T{\mathsf{T}}%
\global\long\def\N{\mathcal{N}}%
\global\long\def\bR{\mathbb{R}}%
\global\long\def\tQ{\tilde{Q}}%
\global\long\def\I{\mathbb{I}}%
\global\long\def\const{\mathrm{const}}%
\global\long\def\tr{\mathrm{tr}}%
\global\long\def\llangle{\langle\!\langle}%
\global\long\def\rrangle{\rangle\!\rangle}%
\icmltitlerunning{Multi-Scale Adaptive Theory of Feature Learning}

\twocolumn[
\icmltitle{From Kernels to Features: A Multi-Scale Adaptive Theory of Feature Learning}

\icmlsetsymbol{equal}{*}

\begin{icmlauthorlist}
\icmlauthor{Noa Rubin}{equal,huji}
\icmlauthor{Kirsten Fischer}{equal,fzj,rwth_phd}
\icmlauthor{Javed Lindner}{equal,fzj,rwth,rwth_cosmo}
\icmlauthor{David Dahmen}{fzj}
\icmlauthor{Inbar Seroussi}{telaviv}
\icmlauthor{Zohar Ringel}{huji}
\icmlauthor{Michael Kr\"amer}{rwth_cosmo}
\icmlauthor{Moritz Helias}{fzj,rwth}
\end{icmlauthorlist}

\icmlaffiliation{fzj}{
Institute for Advanced Simulation (IAS-6), Computational and Systems Neuroscience, Jülich Research Centre, Jülich, Germany}
\icmlaffiliation{rwth}{Department of Physics, RWTH Aachen University, Aachen, Germany}
\icmlaffiliation{rwth_phd}{RWTH Aachen University, Aachen, Germany}
\icmlaffiliation{rwth_cosmo}{Institute for Theoretical Particle Physics and Cosmology, RWTH Aachen University, Aachen, Germany}
\icmlaffiliation{telaviv}{Department of Applied Mathematics, School of Mathematical Sciences, Tel-Aviv University, Tel-Aviv, Israel}
\icmlaffiliation{huji}{The Racah Institute of Physics, The Hebrew University of Jerusalem, Jerusalem, Israel}

\icmlcorrespondingauthor{Noa Rubin}{noa.rubin@mail.huji.ac.il}

\icmlkeywords{Machine Learning Theory}

\vskip 0.3in
]
\printAffiliationsAndNotice{\icmlEqualContribution} 
\begin{abstract}
Feature learning in neural networks is crucial for their expressive
power and inductive biases, motivating various theoretical approaches.
Some approaches describe network behavior after training through a
change in kernel scale from initialization, resulting in a generalization
power comparable to a Gaussian process. Conversely, in other approaches
training results in the adaptation of the kernel to the data, involving
directional changes to the kernel. The relationship and respective
strengths of these two views have so far remained unresolved. This
work presents a theoretical framework of multi-scale adaptive feature
learning bridging these two views. Using methods from statistical
mechanics, we derive analytical expressions for network output statistics
which are valid across scaling regimes and in the continuum between
them. A systematic expansion of the network's probability distribution
reveals that mean-field scaling requires only a saddle-point approximation,
while standard scaling necessitates additional correction terms. Remarkably,
we find across regimes that kernel adaptation can be reduced to an
effective kernel rescaling when predicting the mean network output
in the special case of a linear network. However, for linear and non-linear
networks, the multi-scale adaptive approach captures directional feature
learning effects, providing richer insights than what could be recovered
from a rescaling of the kernel alone.
\end{abstract}

\section{Introduction}

A central phenomenon that is essential for explaining the power
of neural networks (NNs) is feature learning (FL), where networks
learn meaningful high-dimensional representations of the data \cite{Bengio13_1798}.
FL plays an increasingly important role in our ability to understand
and rationalize the behavior of large language models (LLMs). Sparse
autoencoders can extract so called monosemantic features from LLMs
that are given by a superposition of layer activations \cite{Bricken2023};
these features allow interpreting and even altering model behavior
\cite{Templeton2024}. Beyond interpretability, FL is essential for
efficient generalization with finite data, as it enhances informative
directions in the learned representations, reducing the sample complexity
of learning functions of these directions \cite{Abbe21_26989,Paccolat_2021_staircase,Dandi23_1}.
Despite its significance, many open questions remain regarding the
mechanisms underlying the emergence of these feature directions.

A well-characterized case in NN theory is the limit of infinite-width
and finite sample size, where networks behave as Gaussian processes
(GPs) \cite{Mackay2003}, characterized by the neural network Gaussian
process (NNGP) kernel \cite{Neal96,Williams98_1203,Matthews18,Lee18}.
However, the NNGP does not capture FL, which emerges at finite network
width, in the proportional limit, where both network width and sample
size tend to infinity proportionally \cite{Li21_031059}, or in certain
scaling regimes \cite{Yang24}. Hence the NNGP fails to capture the
networks' nuanced internal representations that arise from feature
learning \cite{vanMeegen_24_16689}. Multiple theoretical approaches
have emerged to describe this phenomenon, yet there is no consensus
on how to characterize FL. A common approach is to study the change
of the network kernel, though the existing frameworks differ in their
predictions for this change.

One prominent class of theories, which are commonly referred to as
rescaling theories \cite{Li21_031059,Pacelli23_1497,Baglioni24_027301},
predicts that the average network output and variance can be described
by a rescaled NNGP kernel. Initially developed for linear networks
in the standard scaling regime\footnote{A scaling where readout weight variance scales as $1/\text{width}$.},
this framework surprisingly yields impressively accurate predictions
even in mean-field scaling\footnote{A scaling where readout weight variance scales as $\mathrm{1/\mathrm{\text{width}}^{2}}$.}.
Despite the strong FL in this regime, the average network outputs
can be obtained from an output kernel that is a rescaled NNGP kernel.

\begin{figure}
\centering\vskip 0.2in\includegraphics[width=1\columnwidth]{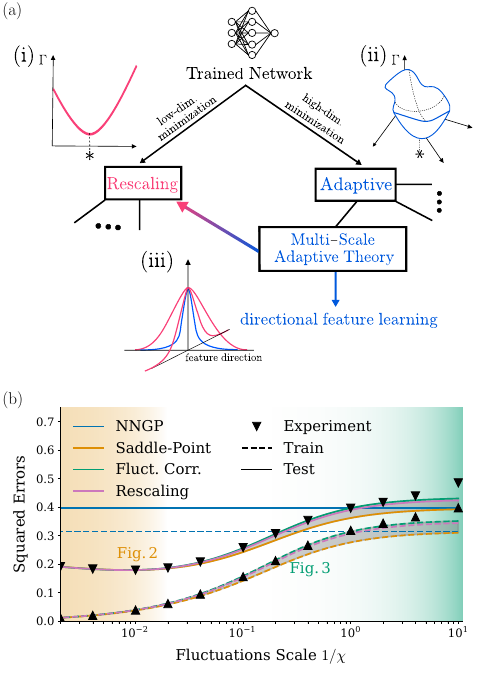}

\caption{(a) The multi-scale adaptive theory bridges between rescaling and
adaptive theories of feature learning. Starting from the distribution
of network outputs for trained networks, the choice of order parameter
decides whether a rescaling (red) or adaptive (blue) theory is obtained.
The choice of order parameter recasts feature learning into either
a (i) low-dimensional minimization or (ii)\textbf{ }high-dimensional
minimization problem. An approximation of the multi-scale adaptive
theory in certain limits yields the result of the rescaling approach,
but in addition describes (iii) directional aspects of feature learning.
(b) Training (solid line) and test errors (dashed line) across scaling
regimes for different approaches. While standard scaling (green shaded
area) requires a one-loop approximation with fluctuation corrections
(Fluct. Corr.), a saddle-point or tree-level approximation (Saddle-Point)
is sufficient in mean-field scaling (orange shaded area). We show
results for the kernel rescaling theory by \cite{Li21_031059} as
reference (Rescaling). We here show results for a linearly separable
task; for results on MNIST see \prettyref{fig:mnist_transition} in
\prettyref{app:add_figures}. Parameters: $\gamma=1$, $P_{\text{train}}=80$,
$N=100$, $D=200$, $\kappa_{0}=1$, $P_{\text{test}}=10^{3}$, $g_{v}=g_{w}=0.5$,
$\Delta p=0.1$.}
\label{fig:graphical_abstract}
\end{figure}

However, FL is often considered a structural phenomenon, as in the
case of Gabor filters \cite{Gabor1946,Rai2020} that emerge in the
latent layers of convolutional neural networks \cite{Luan2018}. Thus,
the expectation is that the effect of FL on the output should be directional
as well. The rescaling result raises fundamental questions about how
learned features are represented in network outputs and can be captured
theoretically.

In contrast, adaptive theories\emph{ }of FL \cite{Roberts22,seroussi23_908,Bordelon23_114009,Fischer24_10761,vanMeegen_24_16689}
consider learned features, predicting that the kernel undergoes a
structural change and incorporates features explicitly. Consequentially,
these theories are able to predict phenomena in networks that stem
from FL such as a reduction in sample complexity -- the required
amount of samples to learn a given task -- relative to that of a
GP \cite{naveh2021a} as well as grokking \cite{Rubin24_iclr}. However,
adaptive theories are significantly more complex computationally than
rescaling theories, while yielding comparable predictions for quantities
such as the network loss. A fundamental open question remains: How
can two such different descriptions of FL be valid at the same time?

In this work, we address this pivotal question by systematically
connecting different FL theories as well as discussing their differences.
Using methods from statistical physics, we recast the theoretical
description of the posterior distribution of network outputs into
a minimization problem with respect to a quantity which we call the
\textquotedblleft order parameter\textquotedblright . We find that
different theories result from different choices of order parameters,
in particular with regard to their dimensionality (see \prettyref{fig:graphical_abstract}a).
Our main contributions are:
\begin{itemize}
\item We derive a multi-scale adaptive theory that is valid across the full
range of scaling regimes, from mean-field to standard scaling, which
allows us to systematically include finite-width corrections (see
\prettyref{fig:graphical_abstract}b). This generalizes previous adaptive
approaches, which were restricted to specific scaling regimes, and
holds for arbitrary tasks as well as non-linear networks.
\item By analyzing the simplest non-trivial model, we reconcile adaptive
and rescaling theories: we show that for the mean network output the
multi-scale adaptive theory can be approximated in certain limits
to yield an effective rescaling of the kernel. This explains why certain
FL phenomena do not appear in rescaling theories.
\item Our theory reveals how the two approaches differ. In linear networks,
rescaling and adaptive theories yield equivalent predictions for the
mean network output, but not for the output covariance. In mean-field
scaling, the covariance exhibits clear adaptation to task-relevant
directions, accurately captured by our adaptive theory and not by
a rescaling theory. In non-linear networks, the disparity between
the approaches emerges already at the level of the mean predictions:
our adaptive theory correctly predicts a change in sample complexity
class relative to the NNGP, a phenomenon that is not captured by rescaling
theories.
\end{itemize}
Overall, our findings suggest that a comprehensive understanding of
FL requires moving beyond kernel rescaling towards high-dimensional
kernel adaptation.

\section{Related works}

The limit of infinite network width and finite amount of training
data has been studied extensively, yielding among others the NNGP
kernel \cite{Neal1995BayesianLF,Williams98_1203,Lee18,Matthews18,avidan2024}.
This theory relates network behavior at initialization to training
dynamics \cite{Poole16_3360,Pennington17_neurips,Schoenholz17_01232,Xiao18_5293}.
However, the NNGP cannot explain the often superior performance of
finite-width networks \cite{Li15_196,Chizat19_neurips,Lee20_ad086f59,Aitchison2020,Refinetti21_8936},
requiring either the inclusion of finite-width effects or different
infinite-width limits such as $\mu$P scaling \cite{Yang22_03466,Vyas23_neurips}.

Describing FL in neural networks in a Bayesian framework has led to
concurrent views: kernel rescaling \cite{Li21_031059,Li22_34789,Pacelli23_1497,bassetti2024,Baglioni24_027301}
and kernel adaptation \cite{Aitchison2020,Naveh21_NeurIPS,seroussi23_908,Fischer24_10761,Rubin24_iclr,vanMeegen_24_16689}.
These differ in the choice of order parameters considered and in consequence
also in the explained phenomena.  A complementary perspective is
provided by \citet{Yang23_pmlr}, who propose a unified theoretical
framework showing that FL extends classical kernel methods by enabling
the learning of data-dependent feature maps in the infinite-width
regime.

Various works study other aspects of networks in the Bayesian framework:
\citet{ZavatoneVeth21_NeurIPS_I} study properties of the network
prior, whereas we focus on the network posterior. \citet{Hanin23}
obtain a rigorous non-asymptotic description of deep linear networks
in terms of Meijer-G functions. \citet{ZavatoneVeth22_064118} study
the same setting but consider explicit models on the input data in
the limit of infinite pattern dimension. \citet{ZavatoneVeth21_600}
investigate deep linear networks in different proportional limits,
recovering the results from Li \& Sompolinsky in an adaptive approach.
\citet{Cui23_6468} study non-linear networks, exploiting the Nishimori
conditions that hold for Bayes-optimal inference, where student and
teacher have the same architecture and the student uses the teacher's
weight distribution as a prior; the latter is assumed Gaussian i.i.d.,
which allows them to use the Gaussian equivalence principle \cite{Goldt20_14709}
to obtain closed-form solutions.

Our work is distinct from perturbative approaches such as \cite{Antognini19_arxiv,Naveh21_064301,Cohen21_023034,Halverson21_035002,Roberts22,Hanin23,Hanin24}
for the Bayesian setting or \cite{Dyer20_ICLR,Huang20_4542,Aitken20_06687,Roberts22,Bordelon23_114009,Buzaglo24_5035}
for gradient-based training that use the strength of non-Gaussian
cumulants of the outputs as an expansion parameter. In contrast, we
perform an expansion in terms of fluctuations around the mean outputs,
which is able to capture phenomena that escape perturbative treatments
such as phase transitions; this technique corresponds to an infinite
resummation of perturbative terms. Our approach is similar to \citet{vanMeegen_24_16689}
with the difference that they scale weight variances as $1/N^{3}$,
so that readout weights concentrate (see \prettyref{app:rel_vanMeegen_somp}
for a comparison of the approaches).

Another line of work focuses on the dynamics of FL: \citet{Saxe14_iclr}
derive exact learning dynamics for deep linear networks, while \citet{Bordelon23_114009}
use dynamical mean-field theory to describe network behavior in the
early stages of training of gradient descent training in different
scaling regimes; in contrast we consider networks at equilibrium.
\citet{Yang20_14522} consider the effect of network training dynamics
and learning rate scales in networks. An experimental investigation
of kernels in feature learning in gradient descent settings was performed
by \citet{Canatar22_ieee}. \citet{day2024} study the effect of
weight initialization on generalization and training speed. A different
viewpoint considers spectral properties of FL \cite{Simon23,Yang24}
as well as investigating the effects of learned representations directly
\cite{Petrini23_114003}. \citet{Maillard24} derive polynomial scaling
limits of the required amount of training data.

\section{Single hidden-layer network\label{sec:setup}}

We consider the following network architecture
\begin{align}
h_{\alpha}=Vx_{\alpha},\;f_{\alpha}=w^{\T}\phi(h_{\alpha}),\;y_{\alpha} & =f_{\alpha}+\xi_{\alpha},\label{eq:network_architecture}
\end{align}
where $\xi$ is Gaussian regularization noise $\xi\stackrel{\text{i.i.d.}}{\sim}\N(0,\kappa)$.
We consider $P$ tuples of training data $\mathcal{D}=\{(x_{\alpha},y_{\alpha})\}_{1\le\alpha\le P}$
with $x_{\alpha}\in\mathbb{\bR}^{D}$ and $y_{\alpha}\in\mathbb{\bR}$
as well as an unseen test point $(x_{*},y_{*})$ denoted by $*$.
Here, $\phi$ denotes a non-linear activation function and $f_{\alpha}\in\bR$
is the scalar network output. We study the Bayesian setting with Gaussian
priors on the readin weights $V\in\bR^{N\times D}$ as $V_{ij}\sim\mathcal{N}(0,g_{v}/D)$
and the readout weights $w\in\bR^{N}$ as $w_{i}\sim\mathcal{N}(0,g_{w}/N^{\gamma})$.

We differentiate between two cases: (a) standard scaling for $\gamma=1$
and (b) mean-field scaling for ${\gamma=2}$. Accordingly, we scale
the regularization noise as $\kappa=\kappa_{0}N^{1-\gamma}$ so that
it does not dominate the network output in mean-field scaling. For
concise notation, we use the shorthands $f_{\mathcal{D}}=(f_{\alpha})_{1\le\alpha\le P}$,
$X=(x_{\alpha})_{1\le\alpha\le P}$ and $y=(y_{\alpha})_{1\le\alpha\le P}$
in the following. Further, summations over repeated indices are implied
$V_{kl}x_{l}\equiv\sum_{l=1}^{N}V_{kl}x_{l}$. The code for theory
and experiments can be found in \citet{Rubin25_zenodo}.

\section{Multi-scale adaptive feature learning theory\label{sec:multi_scale_approach}}

We compute the network posterior on the test point $(x_{*},y_{*})$
by conditioning on the training data $\mathcal{D}$ and derive a set
of self-consistency equations for the average discrepancies $\langle\Delta\rangle$
between labels $y$ and mean posterior network outputs $\langle f_{\mathcal{D}}\rangle$
on the training data. This description on the level of the discrepancies
yields a high-resolution picture of the network behavior: it allows
us to explain kernel rescaling results in the proportional limit as
well as predict directional aspects of FL.
\begin{figure*}[t]
\vskip 0.2in
\begin{centering}
\includegraphics[width=1\textwidth]{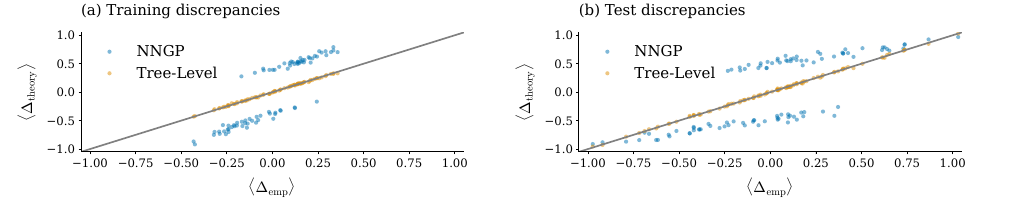}\caption{(a) Training discrepancies $\langle\Delta\rangle=y-\langle f_{\mathcal{D}}\rangle$
and (b) test discrepancies $\langle\Delta_{\ast}\rangle=y_{\ast}-\langle f_{\ast}\rangle$
on an Ising task in mean-field scaling. We show theoretical values
for both NNGP and tree-level against empirical results, where the
gray line marks the identity. In contrast to the NNGP, the tree-level
approximation accurately matches the empirical values. While we use
$\phi=\text{id}$ here, the non-linear case $\phi=\text{erf}$ yields
similar results (see \prettyref{fig:nonlinear_scatter} in \prettyref{app:add_figures}).
Parameters: $\gamma=2$, $P_{\text{train}}=80$, $N=100$, $D=200$,
$\kappa_{0}=1$, $P_{\text{test}}=10^{3}$, $g_{v}=g_{w}=0.5$, $\Delta p=0.1$.}
\label{fig:scatter_mean_field}
\par\end{centering}
\vskip -0.2in
\end{figure*}

\subsection{Predictor statistics of the neural network\label{subsec:predictor_stats}}

We are interested in the Bayesian network posterior for the network
output on training points $f_{\mathcal{D}}$ and a test point $f_{*}$,
which corresponds to training the network with Langevin stochastic
gradient descent (LSGD) \cite{Welling11_681,Stephan17_1,Naveh21_064301}
until convergence (see \prettyref{app:tasks_training} for details).
We denote the joint vector of outputs as ${f\coloneqq(f_{\mathcal{D}},f_{\ast})\in\mathbb{R}^{P+1}}$.
Following along the lines of \citet{Segadlo22_103401}, we write the
joint distribution as
\begin{align}
 & p(f_{\mathcal{D}},f_{\ast},y)=\N(y\vert f_{\mathcal{D}},\kappa_{0}N^{1-\gamma})\label{eq:joint_output}\\
 & \quad\times\int\mathrm{d}\tilde{f}_{\mathcal{D}}\int\mathrm{d}\tilde{f}_{\ast}\,\exp\big[-i\tilde{f}^{\T}f+W(i\tilde{f}_{\mathcal{D}},i\tilde{f}_{\ast})\big],\nonumber 
\end{align}
with $\tilde{f}\coloneqq(\tilde{f}_{\mathcal{D}},\tilde{f_{*}})$
the conjugate fields to ${f=(f_{\mathcal{D}},f_{\ast})}$. The cumulant-generating
function $W(i\tilde{f}_{\mathcal{D}},i\tilde{f_{*}})$ of the network
prior is given by
\begin{align}
W(i\tilde{f}_{\mathcal{D}},i\tilde{f}_{\ast}) & =\ln\left\langle \exp\left(\sum_{a=1}^{P+1}i\tilde{f_{a}}w_{j}\phi\big(h_{aj}\big)\right)\right\rangle _{w_{j},h_{aj}},\label{eq:def_W_main}
\end{align}
where the average $\langle\dots\rangle_{w_{j},h_{j}}$ is over the
prior distribution on the network parameters and the hidden pre-activations
$h_{aj}\overset{\text{i.i.d. over j}}{\sim}\N(0,C^{(xx)})$ with $C^{(xx)}=g_{v}/D\,XX^{\T}\in\bR^{(P+1)\times(P+1)}$.
The detailed derivation can be found in \prettyref{app:theory}. The
statistics of the conjugate fields $\tilde{f}_{\mathcal{D}}$ are
directly linked to the statistics of the network predictors $f_{\mathcal{D}}$
via the output discrepancies ${\Delta=y-f_{\mathcal{D}}}$ on the
training data as
\begin{equation}
\langle\Delta\rangle=-i\kappa_{0}N^{1-\gamma}\langle\tilde{f}_{\mathcal{D}}\rangle.\label{eq:expDelta_exptilf}
\end{equation}
To obtain the statistics of the conjugate variables $(\tilde{f}_{\mathcal{D}},\tilde{f_{*}})$
and thus also of the network outputs $(f_{\mathcal{D}},f_{\ast})$,
we define a conditional cumulant-generating function ${\mathcal{W}(k,j_{\ast}|y)\coloneqq\ln\,\langle\exp(j_{\ast}f_{\ast}+ik^{\T}\tilde{f}_{\mathcal{D}})\rangle_{f_{\ast},\tilde{f}_{\mathcal{D}}}}$
which yields
\begin{align}
\mathcal{W}(k,j_{\ast}\vert y) & =\ln\int\mathrm{d}\tilde{f}\exp\big[ik^{\T}\tilde{f}_{\mathcal{D}}+\mathcal{S}(\tilde{f}_{\mathcal{D}},j_{\ast}\vert y)\big],\label{eq:cum_gen_posterior}\\
\mathcal{S}(\tilde{f}_{\mathcal{D}},j_{\ast}\vert y) & =-iy^{\T}\tilde{f}_{\mathcal{D}}-\frac{\kappa_{0}}{2}N^{1-\gamma}\tilde{f}_{\mathcal{D}}^{\T}\tilde{f}_{\mathcal{D}}\nonumber \\
 & \phantom{=}+W(i\tilde{f}_{\mathcal{D}},j_{\ast}).\nonumber 
\end{align}
Here, we introduced source terms $(k,j_{\ast})$ with $k\in\mathbb{R}^{P}$,
$j_{*}\in\mathbb{R}$ to obtain the statistics of $(\tilde{f}_{\mathcal{D}},f_{*})$
as derivatives. On the training points, we have
\begin{align}
\langle\tilde{f}_{\mathcal{D}}\rangle & =-i\nabla_{k}\mathcal{W}\vert_{k,j_{\ast}=0},\:\llangle\tilde{f}_{\mathcal{D}}\tilde{f}_{\mathcal{D}}^{\T}\rrangle=-\nabla_{k}^{2}\mathcal{W}\vert_{k,j_{\ast}=0},
\end{align}
with $\llangle\tilde{f}_{\mathcal{D}}\tilde{f}_{\mathcal{D}}^{\T}\rrangle$
the covariance of $\tilde{f}_{\mathcal{D}}$. On the test point, we
get
\begin{align}
\langle f_{\ast}\rangle & =\partial_{j_{\ast}}\mathcal{W}\vert_{k,j_{\ast}=0},\;\llangle f_{\ast}^{2}\rrangle=\partial_{j_{\ast}}^{2}\mathcal{W}\vert_{k,j_{\ast}=0}.
\end{align}
However, the cumulant-generating function $\mathcal{W}(k,j_{\ast}\vert y)$
in \prettyref{eq:cum_gen_posterior} in general does not have an analytical
solution. Instead, we perform a systematic expansion in terms of fluctuations
of the network output using the Legendre transform of the cumulant-generating
function $\mathcal{W}$ of the network posterior
\begin{equation}
\Gamma(\bar{\tilde{f}},j_{\ast}\vert y)=\text{extr}_{k}\,ik^{\T}\bar{\tilde{f}}-\mathcal{W}(k,j_{\ast}\vert y),\label{eq:legendre_transform}
\end{equation}
where we take the extremum with respect to $k$. This transform is
a function of the mean conjugate field ${\bar{\tilde{f}}=\langle\tilde{f}_{\mathcal{D}}\rangle}$
(in the following, we drop the index $\mathcal{D}$ for readability),
defined self-consistently by the stationary condition given by
\begin{equation}
\partial_{\bar{\tilde{f}}}\Gamma(\bar{\tilde{f}},j_{\ast}\vert y)\overset{!}{=}0.\label{eq:self_consistent_f_tilde}
\end{equation}
The Legendre transform $\Gamma(\bar{\tilde{f}},j_{\ast}\vert y)$
is thus a natural way of constructing a minimization problem that
yields the quantity we are interested in. It recasts the problem of
computing the statistics of the posterior, which is the stationary
solution of the stochastic minimization problem described by the LSGD
training, into an effective deterministic optimization problem of
$\Gamma$ with regard to the mean discrepancies $\bar{\tilde{f}}$;
intuitively, we may therefore think of $\Gamma$ as an effective loss
function that explicitly depends only on the mean discrepancies $\langle\Delta\rangle\propto\bar{\tilde{f}}$,
but implicitly takes fluctuations of $\Delta$ into account. Moreover,
it allows computing corrections to the mean network outputs in a systematic
manner, building on a broad foundation of methods from statistical
physics \cite{ZinnJustin96,Helias20_970}.

Using the relationship between first-order parametric derivatives
of the Legendre transform $\Gamma$ and the cumulant-generating function
$\mathcal{W}(k,j_{\ast}\vert y)$, we obtain
\begin{equation}
\begin{alignedat}{1}\langle f_{*}\rangle & =-\partial_{j_{\ast}}\Gamma\vert_{k,j^{*}=0}.\end{alignedat}
\label{eq:mean_predictor}
\end{equation}
Next, we consider systematic approximations of $\Gamma$ for different
scaling regimes, from which we determine the network output statistics.

\subsection{Exact network prior for linear networks\label{subsec:exact_linear_prior}}

The above expressions are exact but require knowledge of the cumulant-generating
function $W$ of the network prior in \prettyref{eq:def_W_main}.
For general non-linear activation functions $\phi$, the cumulant-generating
function $W$ can only be approximated, e.g. using a cumulant expansion
(see \prettyref{app:cum_gen_non_lin} for details). For linear activations
$\phi(h)=h$, however, we derive an exact expression
\begin{align}
W(i\tilde{f}_{\mathcal{D}},i\tilde{f}_{\ast}) & =-\frac{N}{2}\ln\det\left[\mathbb{I}+\frac{g_{w}}{N^{\gamma}}\hat{C}^{(xx)}\tilde{f}\,\tilde{f}^{\T}\right]\,.\label{eq:W_linear}
\end{align}
Since the goal of this work is to connect existing theories while
also studying their differences, we consider the linear setting in
the following as the simplest setting possible. Despite this choice,
the theoretical framework presented here also applies to the non-linear
case as discussed in \prettyref{app:theory}.
\begin{figure*}
\vskip 0.2in
\begin{centering}
\includegraphics[width=1\textwidth]{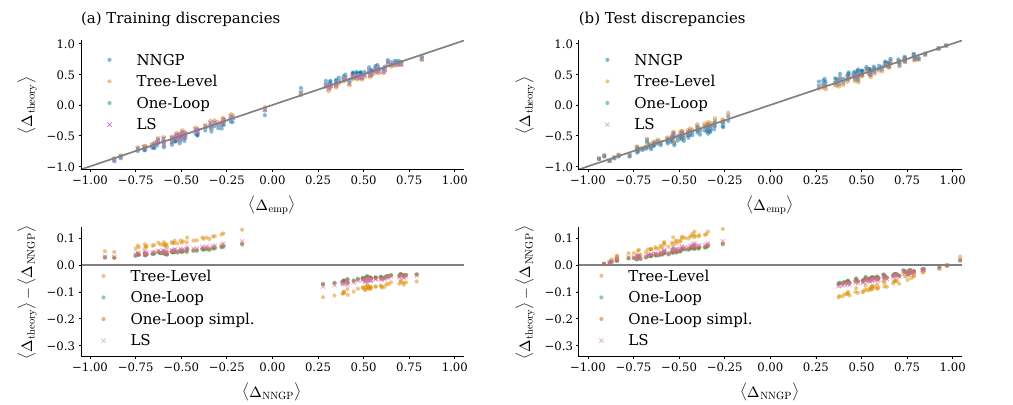}
\par\end{centering}
\begin{centering}
\caption{(a) Training discrepancies $\langle\Delta\rangle=y-\langle f_{\mathcal{D}}\rangle$
and (b) test discrepancies $\langle\Delta_{\ast}\rangle=y_{\ast}-\langle f_{\ast}\rangle$
on an Ising task in standard scaling. Upper row: theoretical values
for different theories against empirical results; gray line marks
the identity. Lower row: difference of theoretical values to the NNGP
as a baseline against NNGP predictions, indicating small differences
between the different approaches. Results of the kernel rescaling
approach by \citet{Li21_031059} are shown as reference (LS). Parameters:
$\gamma=1$, $P_{\text{train}}=80$, $N=100$, $D=200$, $\kappa_{0}=0.4$,
$P_{\text{test}}=10^{3}$, $g_{v}=0.5$, $g_{w}=0.2$, $\Delta p=0.1$.}
\label{fig:scatter_standard_scaling_nngp}
\par\end{centering}
\vskip -0.2in
\end{figure*}

\subsection{Saddle-point approximation in mean-field scaling\label{subsec:mf_scaling}}

In mean-field scaling, the exponent $\mathcal{S}$ of the cumulant-generating
function in \prettyref{eq:cum_gen_posterior} scales linearly with
the network width $N$, while the fluctuations of the network output
scale as $\llangle ff^{\T}\rrangle\sim1/N$ and become negligible.
Thus, we can perform a saddle-point approximation for the integral
in \prettyref{eq:cum_gen_posterior} and obtain the tree-level approximation
of the Legendre transform \cite{Helias20_970} by replacing $\tilde{f}$
by its mean value $\tilde{f}\mapsto\bar{\tilde{f}}$ yielding
\begin{equation}
\Gamma(\bar{\tilde{f}},j_{\ast}\vert y)\approx\Gamma_{\text{TL}}(\bar{\tilde{f}},j_{\ast}\vert y)=-\mathcal{S}(\bar{\tilde{f}},j_{\ast}\vert y)\,.
\end{equation}
 We derive this result more rigorously in \prettyref{app:tree-level_app}
using a large deviation principle \cite{Touchette09}. From the stationary
condition in \prettyref{eq:self_consistent_f_tilde}, we obtain a
self-consistency equation for $\bar{\tilde{f}}$ given by
\begin{align}
\bar{\tilde{f}} & =-i\left(\kappa_{0}N^{-1}\mathbb{I}+C_{\text{TL}}\big(\bar{\tilde{f}}\big)\,C_{\mathcal{D}\mathcal{D}}^{(xx)}\right)^{-1}y\,,\label{eq:self_cons_tl}\\
C_{\text{TL}}\big(\bar{\tilde{f}}\big) & =g_{w}N^{-1}\left[\mathbb{I}+\frac{g_{w}}{N^{2}}C_{\mathcal{D}\mathcal{D}}^{(xx)}\bar{\tilde{f}}\bar{\tilde{f}}^{\T}\right]^{-1},
\end{align}
where $C_{\mathcal{D}\mathcal{D}}^{(xx)}\in\mathbb{R}^{P\times P}$
refers to the training data submatrix of $C^{(xx)}$. In the remainder
of this section, $\bar{\tilde{f}}$ refers to the solution of \prettyref{eq:self_cons_tl}.
We obtain the discrepancies on the training points as
\begin{align}
\langle\Delta\rangle_{\text{TL}} & =\kappa_{0}\left(\kappa_{0}\mathbb{I}+C_{\text{TL}}\big(\bar{\tilde{f}}\big)\,C_{\mathcal{D}\mathcal{D}}^{(xx)}\right)^{-1}y\,.
\end{align}
For the test point, we get
\begin{align}
 & \langle f_{\ast}\rangle_{\text{TL}}\\
 & =\left[C_{\text{TL}}\big(\bar{\tilde{f}}\big)\,C_{\mathcal{D}\ast}^{(xx)}\right]^{\T}\left(\kappa_{0}N^{1-\gamma}\mathbb{I}+C_{\text{TL}}\big(\bar{\tilde{f}}\big)\,C_{\mathcal{D}\mathcal{D}}^{(xx)}\right)^{-1}y\nonumber 
\end{align}
where $C_{\mathcal{D}*}^{(xx)}:=\left\{ g_{v}/D\,x_{\alpha}\cdot x_{*}\right\} _{1\leq\alpha\leq P}\in\mathbb{R}^{P\times1}$,
recovering results by \citet{seroussi23_908}. In \prettyref{fig:scatter_mean_field},
we compare theoretical values for training and test discrepancies
against empirical measurements for linear networks trained on a linearly
separable Ising task (see \prettyref{app:tasks_training} for details).
Comparing to the NNGP as a baseline, we find that, while the NNGP
fails to match network outputs, the multi-scale adaptive theory accurately
predicts the values observed in trained networks.  For similar results
on non-linear networks, see \prettyref{fig:nonlinear_scatter} in
\prettyref{app:add_figures}.

\subsection{Fluctuation corrections in standard scaling\label{subsec:standard_scaling}}

In standard scaling, output fluctuations are not scaled down by
the network width $N$ and instead become non-negligible. To obtain
the leading-order fluctuation corrections, we expand the exponent
$\mathcal{S}$ of the cumulant-generating function in \eqref{eq:cum_gen_posterior}
around its saddle-point $\bar{\tilde{f}}$ to second order as
\begin{equation}
\mathcal{S}\big(\tilde{f}_{\mathcal{D}},j_{\ast}\vert y\big)\approx\mathcal{S}\big(\bar{\tilde{f}},j_{\ast}\big)+\frac{1}{2}\big(\tilde{f}_{\mathcal{D}}-\bar{\tilde{f}}\big)^{\T}\mathcal{S}^{(2)}\big(\tilde{f}_{\mathcal{D}}-\bar{\tilde{f}}\big)\,,
\end{equation}
where $\mathcal{S}^{(2)}$ denotes the Hessian of $\mathcal{S}(\tilde{f},j_{\ast}\vert y)$
with respect to $\tilde{f}$ at the saddle-point $\bar{\tilde{f}}$.
Calculating the Gaussian integral in \prettyref{eq:cum_gen_posterior},
we obtain the one-loop approximation of the Legendre transform \cite{Helias20_970}
as
\begin{align}
\Gamma_{\text{1-Loop}}(\bar{\tilde{f}},j_{\ast}\vert y) & =-\mathcal{S}(\bar{\tilde{f}},j_{\ast})-\frac{1}{2}\log\det(-\mathcal{S}^{(2)})\,.\label{eq:gamma_one_loop}
\end{align}
The self-consistency equation for $\bar{\tilde{f}}$ from the stationary
condition in \prettyref{eq:self_consistent_f_tilde} is then given
by
\begin{equation}
\bar{\tilde{f}}_{\alpha}=\left[A\big(\bar{\tilde{f}}\big)\right]_{\alpha\beta}^{-1}\left[-iy_{\beta}-\frac{1}{2}\left[\mathcal{S}^{(2)}\right]_{\delta\epsilon}^{-1}\mathcal{S}_{\epsilon\delta\beta}^{(3)}\bigg\vert_{j^{*}=0}\right]\,,\label{eq:self_cons_oneloop}
\end{equation}
where $A\big(\bar{\tilde{f}}\big)=\kappa_{0}\mathbb{I}+C_{\text{TL}}\big(\bar{\tilde{f}}\big)\,C_{\mathcal{D}\mathcal{D}}^{(xx)}$
and $\mathcal{S}^{(n)}$ refers to the $n$-th derivative of the exponent
$\mathcal{S}$ with respect to $\tilde{f}$ evaluated at $\bar{\tilde{f}}$
(see \prettyref{app:tree-level_app} for details). In the remainder
of this section, $\bar{\tilde{f}}$ refers to the self-consistent
solution of \prettyref{eq:self_cons_oneloop}, which is not necessarily
the same as the one of \prettyref{eq:self_cons_tl} in the previous
section. This yields for the training discrepancies $\langle\Delta\rangle_{\text{1-Loop}}=i\kappa_{0}\bar{\tilde{f}}$
as in \eqref{eq:expDelta_exptilf} and for the test point from \eqref{eq:mean_predictor}
\begin{align}
 & \langle f_{\ast}\rangle_{\text{1-Loop}}\\
 & =\kappa_{0}^{-1}C_{\ast\mathcal{D}}^{(xx)}C_{\text{TL}}\big(\bar{\tilde{f}}\big)^{\T}\langle\Delta\rangle+\frac{1}{2}(\mathcal{S}^{(2)})_{\beta\alpha}^{-1}\mathcal{S}_{\alpha\beta\ast}^{(3)}\bigg\vert_{j^{*}=0}\,.\nonumber 
\end{align}
In the next section, we will see how these expressions reduce to a
kernel rescaling theory in the proportional limit ${N\propto P\rightarrow\infty}$,
which in linear networks we refer to as one-loop simplified in \prettyref{fig:scatter_standard_scaling_nngp},
where we compare theoretical predictions to empirical measurements
on the Ising task. We show results for the multi-scale adaptive theory
presented here as well as the rescaling theory by \citet{Li21_031059},
which was derived for the standard scaling regime. Due to the weak
FL in standard scaling, all theories match the network behavior relatively
well. However, by taking the NNGP as a reference, the differences
between the theories become discernable: The tree-level solution shows
deviations from the other solutions, predicting overly small test
errors compared to the one-loop solution and compared to empirics.
Furthermore, predictions of the one-loop solution agree to those of
the rescaling theory by \citet{Li21_031059}.

The one-loop solution takes into account leading-order fluctuation
corrections. The latter vanish in mean-field scaling, so one expects
the one-loop approximation to converge to the tree-level result in
this scaling regime. We show this explicitly in \prettyref{fig:graphical_abstract}b,
where we demonstrate how the different theories transition between
the two scaling regimes by scaling $\kappa_{0}\mapsto\kappa_{0}/\chi$
and $g_{w}\mapsto g_{w}/\chi$ with $0.1/N<1/\chi<10$ determining
the scale of fluctuations. As expected, train and test errors decrease
for increasing FL in the mean-field regime. Due to non-negligible
fluctuations, the tree-level and one-loop solutions differ in standard
scaling. When further increasing the fluctuations scale, even the
one-loop solution does not accurately predict empirical measurements
anymore since this regime requires fluctuation corrections beyond
first order. In principle, the multi-scale adaptive approach allows
computing these higher-order correction terms \cite{Helias20_970}.
When decreasing the fluctuations towards the mean-field scaling regime,
the one-loop solution converges to the tree-level solution. Notably,
the here presented multi-scale adaptive approach accurately predicts
train and test errors across both scaling regimes, including the intermediate
regime.

\section{Kernel rescaling theory as an approximation of the multi-scale adaptive
theory\label{sec:connecting_approaches}}

Existing rescaling theories \cite{Li21_031059,Li22_34789,Pacelli23_1497,bassetti2024,Baglioni24_027301}
and adaptive theories \cite{Naveh21_NeurIPS,seroussi23_908,Fischer24_10761,Rubin24_iclr,vanMeegen_24_16689}
make both qualitatively and quantitatively different predictions regarding
network behavior. On the one hand, rescaling approaches predict that
the mean network output is equivalent to that obtained by a rescaled
NNGP kernel. On the other hand, adaptive approaches such as the multi-scale
adaptive theory presented here, as well as other existing approaches,
predict that the kernel adapts to the data in a richer manner, showing
changes in specific directions that are determined by the training
data's statistics. While these approaches are quite different, we
here expose the close relation between them in two respects: (i) We
show that the adaptive and the rescaling approach can both be derived
from the same starting point; the expression for the joint distribution
of the network outputs \prettyref{eq:joint_output}. (ii) We show
that\emph{ }for linear\emph{ }networks the adaptive approach in the
proportional limit ${N\propto P\rightarrow\infty}$ can be approximated
by a kernel rescaling for the mean outputs. For output fluctuations
and for certain non-linear networks, however, such a reduction does
not hold.

Technically, the differences between the two viewpoints stem from
different choices of the order parameter used in the approximation
of the posterior, utilizing either a saddle-point approximation or
including fluctuation corrections. Specifically, with point (i), we
show in \prettyref{app:kernel_scaling_app} that the equations obtained
by \citet{Li21_031059} can be obtained from \prettyref{eq:joint_output}
by marginalizing over the hidden pre-activations $h$ in \eqref{eq:def_W_main}
and performing a change of variables so that the posterior is a function
of a single scalar order parameter $Q:=\|w\|^{2}$. A saddle-point
approximation with respect to this variable yields a self-consistency
equation for $Q$ and consequently expressions for the predictor statistics
on test and training points, such as the mean and fluctuations. As
the order parameter is scalar here, it is limited to describing scalar
changes to the kernel.

Conversely, the choice of the high-dimensional order parameter in
the multi-scale adaptive approach, which in mean-field scaling reproduces
equations from the approach in \cite{seroussi23_908}, results in
structural changes to the kernel. Notably, the choice of a high-dimensional
order parameter results in the need to correct for fluctuations that
arise in standard scaling, requiring us to go beyond the saddle-point
approximation by using fluctuation corrections.

Surprisingly, as we have shown in the previous section, for a linear
network on the level of the mean predictors, the multi-scale adaptive
approach converges to that of the rescaling one, even though they
have qualitatively different kernels. This observation motivates (ii),
showing that for a linear network in the proportional limit ${N\propto P\rightarrow\infty}$,
regardless of the initial choice of order parameter, the mean network
output can be obtained from kernel regression \cite{WilliamsRasmussen06}
with a rescaled NNGP kernel. For non-linear networks, however, differences
already arise on the level of the mean predictors, as shown in the
next section.

In the kernel rescaling case, the predictor for the mean output is
obtained by replacing the NNGP kernel ${K_{\text{NNGP}}=g_{w}N^{1-\gamma}C_{\mathcal{D\mathcal{D}}}^{(xx)}}$
with a rescaled kernel 
\begin{equation}
K_{\text{rescaling}}=Q/(g_{w}N^{1-\gamma})\,K_{\text{NNGP}}\,.
\end{equation}
For the multi-scale adaptive approach presented here, the output statistics
in mean-field scaling are obtained from
\begin{equation}
K_{\text{adaptive, TL }}=\left[\mathbb{I}+\frac{g_{w}}{N^{\gamma}}C_{\mathcal{D\mathcal{D}}}^{(xx)}\bar{\tilde{f}}_{\text{TL}}\bar{\tilde{f}}_{\text{TL}}^{\T}\right]^{-1}K_{\text{NNGP}}\,.
\end{equation}
The appearing matrix product allows a non-trivial change of the NNGP
kernel in certain meaningful directions, e.g. a teacher direction
or dominant eigendirections of the input kernel that align with the
target, thereby yielding additional insights. However, we derive
an equivalent equation for the mean predictor by simplifying \prettyref{eq:W_linear}
using the matrix-determinant-lemma, which yields the mean output from
a rescaled NNGP kernel given by 
\begin{equation}
K_{\text{rescaling, TL }}=Q_{\text{TL}}\big(\bar{\tilde{f}}\big)/(g_{w}N^{1-\gamma})\,K_{\text{NNGP}}\,,
\end{equation}
where $Q_{\text{TL}}\big(\bar{\tilde{f}}\big)=g_{w}N^{1-\gamma}/\big(1+\frac{g_{w}}{N^{\gamma}}\bar{\tilde{f}}^{\T}C^{(xx)}\bar{\tilde{f}}\big)$
and $\bar{\tilde{f}}$ satisfies \prettyref{eq:self_cons_tl}. So
even though the adaptive approach in mean-field scaling considers
a directional change to the kernel, in terms of the mean output this
is equivalent to a rescaled kernel.

In standard scaling, one cannot immediately express the mean output
in terms of a rescaled kernel. However, in the proportional limit
${N\propto P\rightarrow\infty}$, certain fluctuation correction terms
become negligible, reducing the expressions to a rescaling form again
(see \prettyref{app:connection_adpative_kernel_app}). The rescaling
factor is given by
\begin{align}
 & Q_{\text{1-loop}}(\bar{\tilde{f}})\\
 & =Q_{\text{TL}}(\bar{\tilde{f}})-\frac{Q_{\text{TL}}^{2}(\bar{\tilde{f}})}{N}\text{Tr}\left[A^{-1}(\bar{\tilde{f}})\,C_{\mathcal{D\mathcal{D}}}^{(xx)}\right]\,,\nonumber 
\end{align}
where $A(\bar{\tilde{f}})\coloneqq\kappa_{0}\mathbb{I}+Q_{\text{TL}}\big(\bar{\tilde{f}}\big)C_{\mathcal{D\mathcal{D}}}^{(xx)}$,
and $\bar{\tilde{f}}$ satisfies \prettyref{eq:self_cons_eq_oneloop_approx}.

We thus find that known theoretical approaches are all derived from
the same original posterior distribution by considering different
order parameters, while their resulting predictions for the mean network
output behave like a rescaled NNGP. However the rescaling equivalence
of mean predictors holds only for linear networks, as well as non-linear
networks approximated using a cumulant expansion of the non-linearity
as described in \prettyref{app:cum_gen_fun_adapt}, since one only
needs to substitute $C_{\mathcal{D\mathcal{D}}}^{(xx)}\mapsto C_{\mathcal{D\mathcal{D}}}^{(\phi\phi)}$.
However this equivalence does not necessarily hold for more fine-grained
methods of approximating the non-linearity, such as the approach described
in \prettyref{app:VGA}. Applying such approximations within the adaptive
approach allows predicting various phenomena that emerge in non-linear
networks such as phase transitions \cite{Rubin24_iclr} and changes
to sample complexity \prettyref{app:VGA}, that escape a description
by a rescaled kernel.
\begin{figure}
\vskip 0.2in\includegraphics[viewport=0bp 0bp 792bp 612bp,width=1\columnwidth]{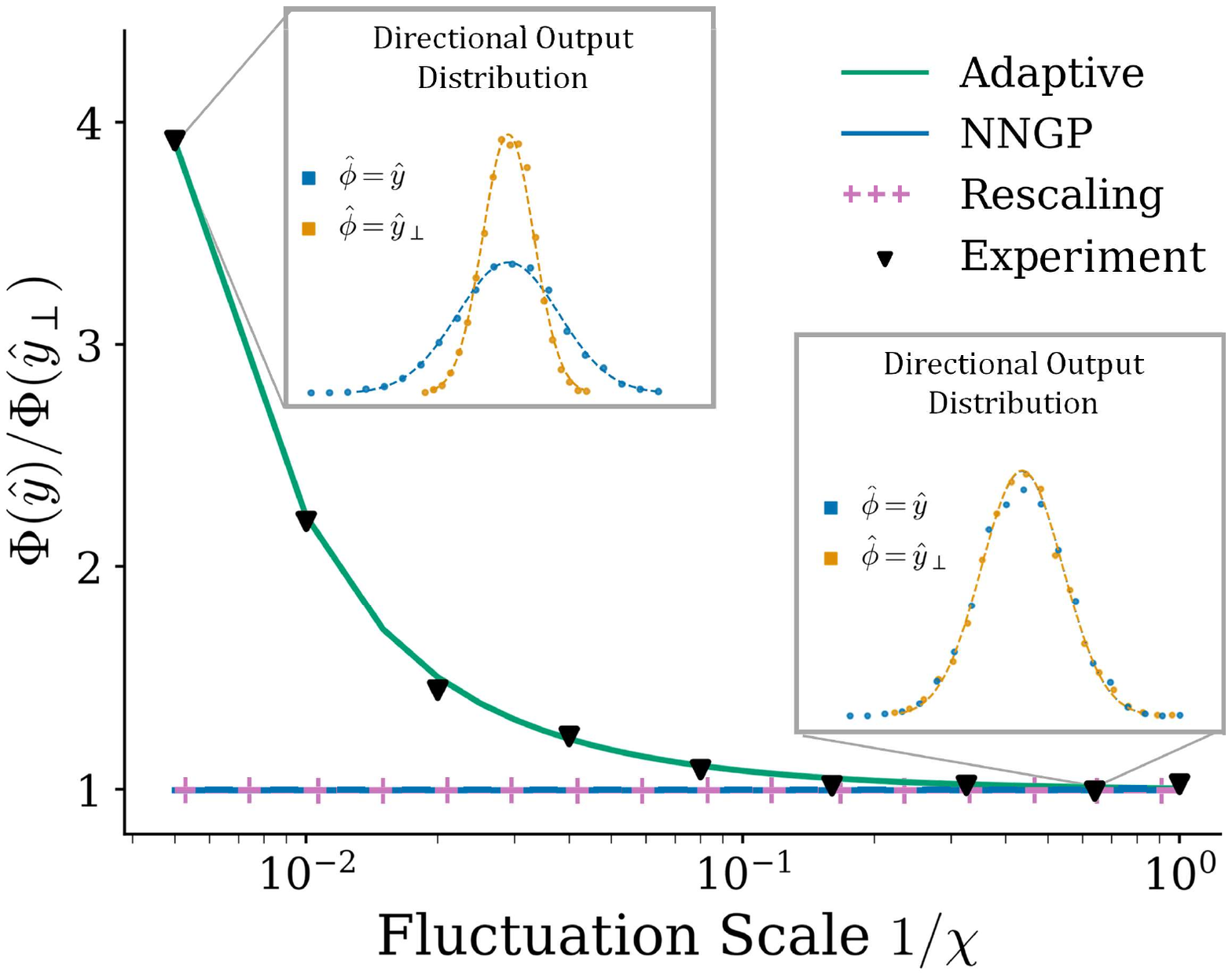}\caption{Relative directional feature learning in a teacher student setting
as a function of the fluctuation scale $1/\chi$. Both NNGP and rescaling
theory fail to capture directional feature learning, while the multi-scale
adaptive theory accurately predicts network behavior. Insets show
the output distribution in different directions; a detailed version
can be found in \prettyref{fig:teacher_student_appendix} in the Appendix.
Parameters: $P_{\text{train}}=80$, $N=200$, $D=50$, $\kappa_{0}=2$,
$g_{v}=0.01,$ $g_{w}=2$.}
\label{fig:fluctuations_adaptation}

\vskip -0.2in
\end{figure}

\section{Directional feature learning emerges in adaptive description\label{sec:directional_fl}}

The power of NNs stems from their ability to detect high-dimensional
features in the data, implying that in the transition from the lazy
to the rich regime this would be reflected in the network output statistics
in a non-trivial manner. It is well established that the network weights
adapt during training in an anisotropic manner, detecting relevant
directions present in the training data \cite{seroussi23_908,Fischer24_10761};
yet surprisingly, for the mean output of a linear network this adaptation
seems to be equivalent to an isotropic rescaling of the NNGP kernel.

\paragraph*{Teacher-student setting}

In this section, we demonstrate that the directional aspect of FL
is nonetheless present in output fluctuations, which is only captured
by the adaptive approach. Given a normalized feature direction $\hat{\phi}$,
we define a directional FL measure $\Phi\big(\hat{\phi}\big)$ that
indicates to which degree this feature is represented by the network
as
\begin{equation}
\Phi\big(\hat{\phi}\big)\coloneqq\hat{\phi}^{\T}\llangle ff^{\T}\rrangle\hat{\phi}.
\end{equation}
Then, $\Phi\big(\hat{\phi}\big)\approx\text{Tr}\left(\llangle ff^{\T}\rrangle\right)$
indicates that the feature direction $\hat{\phi}$ dominates the covariance,
implying that this feature has been perfectly learned, whereas $\Phi\big(\hat{\phi}\big)\ll\text{Tr}\left(\llangle ff^{\T}\rrangle\right)$
is an indication of weak directional FL. As derived in \prettyref{app:tree-level_app},
we obtain for the covariance of the network outputs on the training
data
\begin{align}
\llangle ff^{\T}\rrangle_{\text{adaptive}} & =\kappa\mathbb{I}-\kappa^{2}\big(A+\frac{2Q_{\text{TL}}^{2}}{N\kappa_{0}^{2}}F\big)^{-1},\label{eq:cov_adaptive}
\end{align}
where we observe a structural change in the covariance matrix in form
of the term $F\coloneqq C_{\mathcal{D}\mathcal{D}}^{(xx)}\langle\Delta\rangle\langle\Delta\rangle^{\T}C_{\mathcal{D}\mathcal{D}}^{(xx)}$,
which is not present in a rescaling of the NNGP, whose covariance
is (see \prettyref{eq:cov_rescaling} in \prettyref{app:kernel_scaling_app}
for details)
\begin{equation}
\llangle ff^{\T}\rrangle_{\text{rescaling}}=\kappa\mathbb{I}-\kappa^{2}\big(\kappa\mathbb{I}+Q\,C_{\mathcal{D}\mathcal{D}}^{(xx)}\big)^{-1}.
\end{equation}
As evident from the expressions for the covariance, the directional
FL measure $\Phi$ differs significantly between the two approaches,
which is illustrated most easily for a kernel $C^{(xx)}\propto\mathbb{I}$:
the isotropy of the rescaling theory then results in the same value
of $\Phi\big(\hat{\phi}\big)$ independent of the direction of $\hat{\phi}$,
whereas the structural change of the covariance in the adaptive theory
by the rank-one term $\langle\Delta\rangle\langle\Delta\rangle^{\T}$
in \eqref{eq:cov_adaptive} may yield larger values of $\Phi\big(\hat{\phi}\big)$
for features $\hat{\phi}\parallel\langle\Delta\rangle$.

We show the directional aspect of FL in a teacher-student setting,
where the teacher is given by $y=Xw_{*}$ with $X\sim\mathcal{N}(0,\mathbb{I})$
and the student is a linear network as in \prettyref{sec:setup} with
$\phi(h)=h$. In this setting, the teacher defines a feature direction
$\hat{y}_{*}=Xw_{*}/\left|Xw_{*}\right|$, and for comparison, we
consider another possible feature direction $\hat{y}_{\perp}=Xw_{\perp}/\left|Xw_{\perp}\right|$,
orthogonal to the former in the sense that $w_{\perp}\perp w_{*}$.
The latter can be thought of as the direction of a randomly selected
teacher that differs in the weights of the hidden layer from that
of the actual target teacher. In \prettyref{fig:fluctuations_adaptation},
we show the relative directional FL measure $\Phi\big(\hat{y}_{*}\big)/\Phi\big(\hat{y}_{\perp}\big)$
between the target teacher and a random, orthogonal teacher direction.
While the rescaling theory does not differentiate between these directions,
the adaptive theory accurately predicts amplification of the teacher
direction when entering the mean-field regime.

\paragraph*{Amplification of kernel structure}

The expression \eqref{eq:cov_adaptive} likewise exposes a cooperative
effect caused by feature learning as a result of shaping the kernel
jointly by the input and the output statistics. To illustrate this
effect, we consider the Ising task (see \eqref{app:tasks_training}),
where the input kernel ${C_{\mathcal{D}\mathcal{D}}^{(xx)}=g_{v}\,(\I+\epsilon\,yy^{\T})}$
contains a rank-one term composed of the binary classification labels,
unlike the teacher-student input kernel where $C_{\mathcal{DD}}^{(xx)}\propto\I$.
Although the factor $\epsilon=1+4p(1-p)$ is potentially small, as
is the case for weakly distinguishable patterns ($0\le p-1/2\ll1$),
the matrix factor $F$ appearing in \eqref{eq:cov_adaptive} can significantly
amplify this signal. Consider the case that the target is not learned
well, for example at large ridge $\kappa$, so one has $\langle\Delta\rangle\simeq y$.
The matrix factor
\begin{align}
F & \simeq g_{V}^{2}\,yy^{\T}\,(1+\epsilon\,P)^{2}
\end{align}
then shows that the weak structure $\propto\epsilon$ present in the
input kernel has been amplified by a factor $P$ (see \prettyref{fig:Relative-directional-feature}
in \prettyref{app:add_figures}). For an unstructured kernel $C^{(xx)}\propto\I$
($\epsilon=0$) the rank-one component still persists $F=g_{V}^{2}\,yy^{\T}$,
but it is not amplified by $P$ here. The cooperative effect, in contrast,
shows that feature learning may utilize coherent statistics between
input and output.

\paragraph{Non-linear networks}

Directional effects in linear networks emerge only in the covariance
of the outputs. For non-linear networks, however, these effects appear
in the mean outputs, and have qualitative effects on network performance.
The adaptive approach described in \prettyref{app:VGA} enables structural
changes in the kernel which enhance specific features of the data.
This enhancement, in turn, allows the network to learn non-linear
functions of those features with significantly fewer data points than
would typically be expected. In particular, introducing even a small
non-linearity can lead to a fundamental difference in the behavior
of approaches that rely on a rescaled or adapted kernel. In\prettyref{app:VGA}
we demonstrate this for a teacher-student setting similar to the one
above but where we take both teacher and student to be non-linear.
We find that the adaptive approach accurately predicts that the student
will learn the non-linear component of the teacher, in contrast to
the NNGP and kernel rescaling approaches which predict the non-linearity
would not be learned (see \prettyref{fig:complexity_change_erf} in
\prettyref{app:VGA}). These results reflect a change to the sample
complexity class, and establish a principled connection between feature
learning and improved network performance.

\section{Discussion}

In this work we present a unified theoretical framework to understand
feature learning (FL) in the Bayesian setting across scaling regimes,
from lazy to rich learning. This framework describes both effects
of data adaptation in trained networks, i.e. directional changes of
the network's output statistics in response to statistical dependencies
present in the training data, as well as output rescaling phenomena
that were described in previous works \cite{Li21_031059,Pacelli23_1497}.
Our theory thus creates links between existing and so far unconnected
previous theories. In the rich regime, the presented multi-scale adaptive
theory clearly exposes directional aspects of FL, thus going beyond
rescaling theories. By considering the simplest non-trivial case of
a linear network, we finally reconcile the apparent contradiction
between directional adaptation and rescaling by recovering the latter
as an approximation of the former on the level of the mean output.

Furthermore, the multi-scale adaptive theory presented here applies
to both standard and mean-field scaling and the entirety of the scaling
spectrum. The latter is possible since the presented theoretical framework
allows systematically computing fluctuation corrections depending
on the scaling regime. While the tree-level solution and its equivalence
to rescaling applies only to $\gamma=2$, the one-loop theory applies
to the full regime $\gamma\in[1,2]$ and thus also the equivalence
to rescaling (see \prettyref{sec:connecting_approaches}). Note that
this equivalence is restricted to the mean predictor, while the adaptive
theory captures additional aspects, like directional FL in the variance
terms (see \prettyref{sec:directional_fl}). Moreover, this equivalence
breaks down in the case of non-linear networks, where the presence
of FL results in changes to the sample complexity class. In addition,
our theory does not make any assumptions on the data set; we show
results for an Ising task, a teacher student task and MNIST.

\textbf{Outlook} \;In this work, we study shallow networks, but we
expect directional FL to be crucial for network performance in deeper
networks as well. Beyond this, it will be valuable to extend the theoretical
framework to other network architectures such as convolutional networks,
residual networks, and transformers, using the respective network
priors \cite{Garriga19,Hron20_4376,Fischer24_arxiv}. To study the
effect of noise in input data on FL \cite{Lindner23_arxiv}, we would
like to include fluctuations of the input kernel in the theoretical
framework.

\section*{Acknowledgements}

This work was partly funded by the Deutsche Forschungsgemeinschaft
(DFG, German Research Foundation) - 368482240/GRK2416, and the Israeli
Science Foundation - 374/23. MK would like to thank the Institute
for Advanced Simulation (IAS-6) at Juelich Research Center and its
directors Markus Diesmann and Sonja Gr\"un for their hospitality
during regular visits. We thank the reviewers for providing valuable
feedback.

\section*{Impact Statement}

This paper works towards understanding feature learning, thus aiming
to advance explainability of networks. While the latter surely has
societal impacts, these will be much further down the line.

\bibliographystyle{icml2025_style/icml2025}
\bibliography{brain,add_to_brain}

\newpage{}

\appendix
\onecolumn 

\section{General approach to train and test statistics\label{app:theory}}

We are interested in the training discrepancies $\langle\Delta_{\alpha}\rangle=y_{\alpha}-\langle f_{\alpha}\rangle$
with $\langle f_{\alpha}\rangle$ denoting the mean network output,
and in the mean network output $\langle f_{\ast}\rangle$ for a test
point $x_{\ast}$ after conditioning on the training data $\mathcal{D}=\{(x_{\alpha},y_{\alpha})\}_{1\le\alpha\le P}$.
For clarity, in the appendix we make all index notations explicit
instead of using $\mathcal{D}$ as in the main text, and denote summations
over training data points with Greek letters. We refer to the mean
network outputs as predictors. The joint prior distribution for $(f,f_{\ast},y)$
can be computed as in \cite{Segadlo22_accepted,Fischer24_10761} and
is given by
\begin{align}
p(f,f_{\ast},y) & =\N(y\vert f,\kappa_{0}N^{1-\gamma})\,\int d^{(P)}\tilde{f}\int\text{d}\tilde{f}_{\ast}\exp\bigg(-\sum_{a=1}^{P+1}i\tilde{f}_{a}f_{a}+W(i\tilde{f},i\tilde{f}_{\ast})\bigg),\label{eq:dist_output}\\
W(\tilde{f}_{\mathcal{D}},\tilde{f}_{\ast}) & =\ln\bigg\langle\exp\bigg(\sum_{a=1}^{P+1}\tilde{f_{a}}\sum_{j=1}^{N}w_{j}\,\phi(h_{aj})\bigg)\bigg\rangle_{w_{i},h_{ai}},\label{eq:cum_gen_output}
\end{align}
where we use the shorthands $\int\text{d}^{(P)}\tilde{f}=\prod_{\alpha=1}^{P}\int_{-\infty}^{\infty}d\tilde{f}_{\alpha}/(2\pi)$,
the $P+1$ index corresponds to the test point, and $i$ is the imaginary
unit. The i.i.d. distribution of the readin weights $V_{kl}$ implies
that $h_{\alpha j}\overset{\text{i.i.d. over j}}{\sim}\N(0,\hat{C}^{(xx)})$
with the covariance matrix of the hidden-layer representation given
by
\begin{equation}
\hat{C}^{\left(xx\right)}=\begin{bmatrix}C^{\left(xx\right)} & \left\{ C_{\alpha*}^{(xx)}\right\} _{\alpha=1}^{P}\\
\left\{ C_{*\alpha}^{(xx)}\right\} _{\alpha=1}^{P} & C_{**}^{\left(xx\right)}
\end{bmatrix}\,,\label{eq:cov_matrix_h}
\end{equation}
where $C^{(xx)}=g_{v}/D\,XX^{\T},$and $C_{*\alpha}^{(xx)}=g_{v}/D\ x_{\alpha}\cdot x_{*}$.
To keep notation concise, summations over repeated indices on the
right are implied in the following.

We may obtain training discrepancies $\langle\Delta_{\alpha}\rangle$
and the test predictor $\langle f_{\ast}\rangle$ from the joint cumulant-generating
function $\mathcal{W}$ for the test point defined as
\begin{align}
\mathcal{W}(j_{\ast}\vert y) & =\ln\int df_{\ast}\int df\exp(j_{\ast}f_{\ast})\,p(f,f_{\ast},y).\label{eq:def_calW}
\end{align}
Taking its derivatives w.r.t. to either training labels $y_{\alpha}$
or the source term $j_{\ast}$ yields the posterior of the desired
quantities
\begin{align}
\langle\Delta_{\alpha}\rangle & =-\kappa_{0}N^{1-\gamma}\frac{\partial\mathcal{W}(j_{\ast}\vert y)}{\partial y_{\alpha}}\vert_{j_{\ast}},\label{eq:mean_discrepancy}\\
\langle f_{\ast}\rangle & =\frac{\partial\mathcal{W}(j_{\ast}\vert y)}{\partial j_{\ast}}\bigg\vert_{j_{\ast}=0},
\end{align}
because the outer derivative of the logarithm produces the normalization
by the model evidence (marginal likelihood) $1/p(y)=1/\int df\,\int df_{\ast}\,p(f,f^{\ast},y)$.

Likewise, the variances follow as
\begin{align}
\llangle\Delta_{\alpha}\Delta_{\beta}\rrangle & =\kappa_{0}N^{1-\gamma}-\kappa_{0}^{2}N^{2-2\gamma}\frac{\partial^{2}\mathcal{W}(j_{\ast}\vert y)}{\partial y_{\alpha}\partial y_{\beta}}\vert_{j_{\ast}=0},\label{eq:Var_discrepancy}\\
\llangle f_{*}^{2}\rrangle & =\frac{\partial^{2}\mathcal{W}(j_{\ast}\vert y)}{\partial(j_{\ast})^{2}}\bigg\vert_{j_{\ast}=0},
\end{align}
By inserting \eqref{eq:dist_output} into \eqref{eq:def_calW} and
performing the integration over $f$, we can rewrite $\mathcal{W}$
as
\begin{equation}
\mathcal{W}(j_{\ast}\vert y)=\ln\int d^{(P)}\tilde{f}\,\exp\big(-iy_{\alpha}\tilde{f}_{\alpha}-\frac{\kappa_{0}}{2}N^{1-\gamma}\tilde{f}_{\alpha}\tilde{f}_{\alpha}+W(i\tilde{f}_{\mathcal{D}},j_{\ast})\big).\label{eq:calW_f_integrated}
\end{equation}
Comparing \eqref{eq:mean_discrepancy}, \eqref{eq:Var_discrepancy},
and \eqref{eq:calW_f_integrated}, we note that $y$ acts as a linear
source term for $\tilde{f}_{\mathcal{D}}$, from which we see that
the physical meaning of the field $\tilde{f}_{\mathcal{D}}$ is related
to the discrepancy between target and network output
\begin{align}
\langle\Delta_{\alpha}\rangle & =\kappa_{0}N^{1-\gamma}\langle i\tilde{f}_{\alpha}\rangle,\label{eq:relation_Delta_tilde_f_stats}\\
\llangle\Delta_{\alpha}\Delta_{\beta}\rrangle & =\kappa_{0}N^{1-\gamma}\delta_{\alpha\beta}+\kappa_{0}^{2}N^{2-2\gamma}\,\llangle\tilde{f}_{\alpha}\tilde{f}_{\beta}\rrangle.\label{eq:cov_general}
\end{align}
For computational convenience, we now introduce a source term $k$
\begin{equation}
\mathcal{W}(k,j_{\ast}\vert y)=\ln\int\text{d}\tilde{f}\exp\Big(ik_{\alpha}\tilde{f}_{\alpha}\underbrace{-iy_{\alpha}\tilde{f}_{\alpha}-\frac{\kappa_{0}}{2}N^{1-\gamma}\tilde{f}_{\alpha}\tilde{f}_{\alpha}+W(i\tilde{f},j_{\ast})}_{\mathcal{S}}\Big),\label{eq:cum_gen_full}
\end{equation}
allowing us to compute moments of $\tilde{f}$ by differentiating
by $k$ instead of $y$ and subsequently setting $k=0$. We define
the latter part of the exponent of $\mathcal{W}$ as the action
\begin{equation}
\mathcal{S}(\tilde{f},j_{\ast}\vert y)\coloneqq-iy_{\alpha}\tilde{f}_{\alpha}-\frac{\kappa_{0}}{2}N^{1-\gamma}\tilde{f}_{\alpha}\tilde{f}_{\alpha}+W(i\tilde{f},j_{\ast}\vert y).\label{eq:appendix_action}
\end{equation}

Depending on the scaling in $\gamma$, the network outputs $f$ fully
concentrate on their mean values or require corrections due to non-negligible
fluctuations. To treat both cases jointly and systematically, we introduce
the so-called effective action \cite{Helias20_970} as
\begin{equation}
\Gamma(\bar{\tilde{f}},j_{\ast}\vert y)=\text{extr}_{k}\,ik^{\T}\bar{\tilde{f}}-\mathcal{W}(k,j_{\ast}\vert y),\label{eq:Appendix_VGF_Definition}
\end{equation}
where we explicitly keep the dependence on the source term $j_{\ast}$
for the test point in order to compute parametric derivatives to obtain
test point statistics. This corresponds to the Legendre transform
of the cumulant-generating function $\mathcal{W}$; in the case that
$\mathcal{W}(k,j_{\ast}\vert y)$ has a scaling form, a large deviation
principle can be applied and the effective action corresponds to the
rate function \cite{Touchette09}.

The argument $\bar{\tilde{f}}$ is implicitly defined by the stationary
point (sometimes referred to as the equation of state)
\begin{equation}
\frac{\partial\Gamma(\bar{\tilde{f}},j_{\ast}\vert y)}{\partial\bar{\tilde{f}}_{\alpha}}=ik_{\alpha}=0,\label{eq:appendix_equation_of_state}
\end{equation}
as we set the source term $k$ to $0$ by definition. Using the definition
of $\Gamma$ in \prettyref{eq:Appendix_VGF_Definition}, the extremum
condition yields a self-consistency equation for $\bar{\tilde{f}}$
\begin{equation}
i\bar{\tilde{f}}(j_{\ast})=\frac{\partial\mathcal{W}(k,j_{\ast}\vert y)}{\partial k}.
\end{equation}
In the following we determine approximations of the Legendre transform
$\Gamma(\bar{\tilde{f}},j_{\ast}\vert y)$ to different orders of
statistical fluctuations, corresponding to different scaling regimes.
From the definition of the effective action $\Gamma$ follows as well
that we obtain the mean output on that test point from
\begin{align}
\langle f_{\ast}\rangle & =\frac{\partial\mathcal{W}(k,j_{\ast}\vert y)}{\partial j_{\ast}}\bigg\vert_{k,j_{\ast}=0}=-\frac{\partial\Gamma(\bar{\tilde{f}},j_{\ast}\vert y)}{\partial j_{\ast}}\bigg\vert_{j^{*}=0}.
\end{align}

\subsection{Cumulant-generating function $W$ of the network prior\label{app:cum_gen_fun_adapt}}

We recall that $h_{\alpha j}$ is i.i.d Gaussian in $j$ with covariance
given by \eqref{eq:cov_matrix_h}. Likewise, the weights $w_{i}$
are i.i.d., so that we may factorize the expectation to get an overall
factor of $N$ in \eqref{eq:cum_gen_output}
\begin{align}
W(\tilde{f}_{\mathcal{D}},\tilde{f}_{\ast}) & =N\,\ln\bigg\langle\exp\bigg(\sum_{a=1}^{P+1}\tilde{f_{a}}w\,\phi(h_{a})\bigg)\bigg\rangle_{w,h_{a}},\label{eq:cum_gen_scaling_form}
\end{align}
reducing the expectation over $w$ and $h$ to scalars with regard
to the former neuron index $j$.

\subsubsection{Linear activation function\label{app:cum_gen_linear}}

For linear activation function $\phi(h)=h$, we compute first the
Gaussian integral in \eqref{eq:cum_gen_scaling_form} over $w$ and
then over the hidden-layer representations $h$, yielding
\begin{align}
W(\tilde{f}_{\mathcal{D}},\tilde{f}_{\ast}) & =N\,\ln\bigg\langle\exp\bigg(\tilde{f}^{\T}h\,w\,\bigg)\bigg\rangle_{w,h_{a}}\nonumber \\
 & =-\frac{N}{2}\,\ln\det\left[\mathbb{I}-\frac{g_{w}}{N^{\gamma}}\hat{C}^{(xx)}\tilde{f}\tilde{f}^{\T}\right].\label{eq:cum_gen_linear}
\end{align}
Taking the integrals in the opposite order yields
\begin{align}
W(\tilde{f}_{\mathcal{D}},\tilde{f}_{\ast}) & =N\,\ln\,\Big\langle\exp\Big(\frac{1}{2}\tilde{f}^{\T}\,\hat{C}^{(xx)}\,\tilde{f}\,w^{2}\,\Big)\Big\rangle_{w}\nonumber \\
 & =-\frac{N}{2}\,\ln\left[1-\frac{g_{w}}{N^{\gamma}}\tilde{f}^{\T}\hat{C}^{(xx)}\tilde{f}\right]\,.\label{eq:W_Gauss}
\end{align}
The two expressions are identical also by the matrix-determinant lemma.

\subsubsection{Non-linear activation function\label{app:cum_gen_non_lin}}

For a non-linear activation function $\phi(h)$, the integral in \eqref{eq:cum_gen_scaling_form}
in general cannot be solved in a closed form. However, one may perform
a cumulant-expansion of the cumulant-generating function \eqref{eq:cum_gen_scaling_form}
in terms of the first two cumulants of $\phi$ as
\begin{align}
W(\tilde{f}_{\mathcal{D}},\tilde{f}_{\ast}) & =N\,\ln\,\bigg\langle\exp\bigg(\tilde{f}^{\T}\phi(h)w\,\bigg)\bigg\rangle_{w,h_{a}},\label{eq:W_non_lin}\\
 & =N\,\ln\,\Big\langle\exp\Big(w\,\tilde{f}^{\T}m+\frac{1}{2}w^{2}\,\tilde{f}^{\T}\hat{C}^{(\phi\phi)}\tilde{f}\,\Big)\Big\rangle_{w}\,,\nonumber 
\end{align}
where we introduced the short hands for the cumulants of $\phi$ as
\begin{align}
m_{a} & :=\langle\phi(h_{a})\rangle_{h_{a}\sim\N(0,\hat{C}^{(xx)})}\,,\label{eq:cum_phi}\\
\hat{C}_{ab}^{(\phi\phi)} & :=\big\langle\phi(h_{a})\phi(h_{b})\big\rangle_{(h_{a},h_{b})\sim\N(0,\hat{C}^{(xx)})}\,.\nonumber 
\end{align}
Taking the expectation over $w$ of \eqref{eq:W_non_lin} yields
\begin{align}
W(\tilde{f}_{\mathcal{D}},\tilde{f}_{\ast}) & =-\frac{N}{2}\,\ln\left[1-\frac{g_{w}}{N^{\gamma}}\tilde{f}^{\T}\hat{C}^{(\phi\phi)}\tilde{f}\right]\label{eq:W_nonlin_final}\\
 & +\frac{N}{2}\,\big[\tilde{f}^{\T}m\big]^{2}\,\left[\frac{N^{\gamma}}{g_{w}}-\tilde{f}^{\T}\hat{C}^{(\phi\phi)}\tilde{f}\right]^{-1}\,.\nonumber 
\end{align}
For point-symmetric activation functions $\phi(-h)=-\phi(h)$, such
as erf or tanh activation, the mean $m\equiv0$ vanishes and comparing
to \eqref{eq:W_Gauss} the only replacement that appears is $\hat{C}^{(xx)}\to\hat{C}^{(\phi\phi)}$.

\subsection{Tree-level approximation\label{app:tree-level_app}}

To compute the output statistics, one technically requires the exact
effective action $\Gamma$ in \prettyref{eq:Appendix_VGF_Definition}.
However, in general it does not have an analytical solution and we
instead determine a systematic expansion. A well-established method
from both statistical physics and quantum field theory is the loopwise
expansion \cite{Helias20_970}, expands the effective action $\Gamma(\bar{\tilde{f}})$
in terms of fluctuations of $\tilde{f}$ around its mean value $\bar{\tilde{f}}$.
The lowest-order term of the loopwise expansion is called the tree-level
approximation, which hence corresponds to a standard mean-field approximation:
one replaces $\tilde{f}_{\mathcal{D}}$ by its mean $\bar{\tilde{f}}$in
the action itself
\begin{align}
\Gamma_{\text{TL}}(\bar{\tilde{f}},j_{\ast}\vert y) & =-S(\bar{\tilde{f}},j_{\ast}\vert y)\\
 & =iy_{\alpha}\bar{\tilde{f}}_{\alpha}+\frac{\kappa_{0}}{2}N^{1-\gamma}\bar{\tilde{f}}_{\alpha}\bar{\tilde{f}}_{\alpha}-W(i\bar{\tilde{f}},j_{\ast}).\label{eq:Appendix_Tree_level_VGF}
\end{align}
The average value of $\bar{\tilde{f}}_{\alpha}$ is given by the equation
of state \eqref{eq:appendix_equation_of_state} of the effective action
\begin{equation}
\frac{\partial\Gamma_{\text{TL}}(\bar{\tilde{f}},j_{\ast}\vert y)}{\partial\bar{\tilde{f}}_{\alpha}}\bigg\vert_{j_{\ast}=0}=0.\label{eq:eq_state_tilde_f}
\end{equation}
In mean-field scaling ($\gamma=2$) and for $N\rightarrow\infty$
this result becomes exact using the G\"artner-Ellis theoreom: the
output cumulant-generating function $\mathcal{W}$ in \prettyref{eq:cum_gen_posterior}
has a scaling form as 
\begin{equation}
iy_{\alpha}\tilde{f}_{\alpha}+\frac{\kappa_{0}}{2N}\tilde{f}_{\alpha}\tilde{f}_{\alpha}-W(i\tilde{f}_{\mathcal{D}},j_{\ast})=N\lambda_{f}(\tilde{f}_{\mathcal{D}}/N)
\end{equation}
with $\lambda_{f}(k)=iy_{\alpha}k_{\alpha}+\frac{\kappa_{0}}{2}k_{\alpha}k_{\alpha}-W(ik,j_{\ast})$.
Thus, we can approximate the probability distribution of network outputs
as \cite{Touchette09}
\begin{equation}
-p(y\vert C^{(xx)})/N\approx\Gamma_{\text{TL}}(\bar{\tilde{f}},j_{\ast}\vert y).
\end{equation}
Due to the strong suppression of fluctuations in mean-field scaling
with $N\rightarrow\infty$, the tree-level approximation is sufficient
to describe the network behavior and in particular
\begin{align}
\lim_{N\rightarrow\infty}-p(y\vert C^{(xx)})/N & =\Gamma_{\mathrm{TL}}(\bar{\tilde{f}},j_{\ast}\vert y).\label{eq:appendix_Large_deviation_principle}
\end{align}
However, in the case of larger output fluctuations as in standard
scaling ($\gamma=1$), we need to take into account the output fluctuations
systematically by including higher-order corrections to the tree-level
result. We derive the leading-order correction in the following section
\ref{app:one_loop_app}.

\subsubsection{Linear activation function}

From the equation of state \eqref{eq:eq_state_tilde_f} we obtain
a self-consistency equation for $\bar{\tilde{f}}_{\alpha}$ as
\begin{align}
\bar{\tilde{f}} & =-i\left(\kappa_{0}N^{1-\gamma}\mathbb{I}+C_{\text{TL}}\big(\bar{\tilde{f}}\big)C^{\left(xx\right)}\right)^{-1}y,\\
C_{\text{TL}}\big(\bar{\tilde{f}}\big) & =g_{w}N^{1-\gamma}\left[\mathbb{I}+\frac{g_{w}}{N^{\gamma}}C^{(xx)}\bar{\tilde{f}}\bar{\tilde{f}}^{\T}\right]^{-1}.
\end{align}
Using the relation between the statistics of the discrepancies $\Delta$
and $\tilde{f}_{\mathcal{D}}$ \eqref{eq:relation_Delta_tilde_f_stats},
we obtain for the training discrepancies
\begin{align}
\langle\Delta_{\alpha}\rangle & =-\kappa_{0}N^{1-\gamma}\frac{\partial\mathcal{W}(k,j_{\ast}\vert y)}{\partial k_{\alpha}}\vert_{j_{\ast}=0}\label{eq:Delta_tree_level}\\
 & =\kappa_{0}\left(\kappa_{0}\mathbb{I}+C_{\text{TL}}\big(\bar{\tilde{f}}\big)C^{(xx)}\right)_{\alpha\beta}^{-1}y_{\beta}.
\end{align}
For the test point, we get
\begin{align}
\langle f_{\ast}\rangle_{\text{TL}} & =-\frac{\partial\Gamma_{\text{TL}}(\bar{\tilde{f}},j_{\ast}\vert y)}{\partial j_{\ast}}\rvert_{j_{\ast}=0}=\frac{\partial W(i\bar{\tilde{f}},j_{\ast})}{\partial j_{\ast}}\rvert_{j_{\ast}=0}\label{eq:f_tilde_tree_level}\\
 & =\frac{g_{w}}{N^{1-\gamma}}C_{*\alpha}^{(xx)}\left(\mathbb{I}+\frac{g_{w}}{N^{\gamma}}C^{(xx)}\bar{\tilde{f}}\bar{\tilde{f}}^{\T}\right)_{\alpha\beta}^{-1}i\bar{\tilde{f}}_{\beta}.
\end{align}
By substituting the self-consistency equation for $\bar{\tilde{f}}$,
we obtain
\begin{align}
\langle f_{*}\rangle_{\text{TL}} & =C_{\text{TL}}\big(\bar{\tilde{f}}\big)_{\delta\alpha}C_{*\delta}^{(xx)}\left[\left(\kappa_{0}N^{1-\gamma}\mathbb{I}+C_{\text{TL}}\big(\bar{\tilde{f}}\big)\right)^{-1}\right]_{\alpha\beta}y_{\beta}\,.
\end{align}
For the covariance of the network predictors, we use that $\Delta_{\alpha}=y_{\alpha}-f_{\alpha}$
implies
\begin{align}
\langle\langle\Delta_{\alpha}\Delta_{\beta}\rangle\rangle & =\langle(\Delta_{\alpha}-\langle\Delta_{\alpha}\rangle)(\Delta_{\beta}-\langle\Delta_{\beta}\rangle)\rangle\nonumber \\
 & =\langle(f_{\alpha}-\langle f_{\alpha}\rangle)(f_{\beta}-\langle f_{\beta})\rangle=\langle\langle f_{\alpha}f_{\beta}\rangle\rangle,
\end{align}
so that we may obtain the covariance from \eqref{eq:cov_general}
as
\begin{equation}
\langle\langle f_{\alpha}f_{\beta}\rangle\rangle=\kappa_{0}N^{1-\gamma}\delta_{\alpha\beta}-\kappa_{0}^{2}N^{2-2\gamma}\frac{\partial^{2}\mathcal{W}(j_{\ast}\vert y)}{\partial y_{\alpha}\partial y_{\beta}}\vert_{j_{\ast}=0}.
\end{equation}
Using the involutive property of the Legendre transform, we can express
the Hessian of $\mathcal{W}$ by the inverse of the Hessian of $\Gamma$
using $\mathcal{W}^{(2)}=-(\Gamma^{(2)})^{-1}$, yielding

\begin{equation}
\langle\langle f_{\alpha}f_{\beta}\rangle\rangle=\kappa_{0}N^{1-\gamma}\delta_{\alpha\beta}+\kappa_{0}^{2}N^{2-2\gamma}\left(\frac{\partial^{2}\Gamma(\bar{\tilde{f}}\vert y)}{\partial\bar{\tilde{f}}\partial\bar{\tilde{f}}}\vert_{j_{\ast}=0}\right)^{-1}.
\end{equation}
In tree level approximation with $\Gamma_{\mathrm{TL}}=-\mathcal{S}$,
we then have
\begin{equation}
\langle\langle f_{\alpha}f_{\beta}\rangle\rangle=\kappa_{0}N^{1-\gamma}\delta_{\alpha\beta}+\kappa_{0}^{2}N^{2-2\gamma}\left(\frac{\partial^{2}(-\mathcal{S}(\bar{\tilde{f}}\vert y))}{\partial\overline{\tilde{f}}\partial\overline{\tilde{f}}}\vert_{j_{\ast}=0}\right)^{-1}.
\end{equation}
Computing the Hessian of the action $\mathcal{S}$, we get
\begin{align}
\frac{\partial^{2}(-\mathcal{S}(\bar{\tilde{f}},j_{\ast}))}{\partial\bar{\tilde{f}}_{\alpha}\partial\bar{\tilde{f}}_{\beta}}\bigg\vert_{j^{*}=0} & =\kappa_{0}\delta_{\alpha\beta}+Q_{\text{TL}}\big(\bar{\tilde{f}}\big)\,C_{\alpha\beta}-\frac{2}{N}Q_{\text{TL}}\big(\bar{\tilde{f}}\big)\,C_{\alpha\delta}\bar{\tilde{f}}_{\delta}\,Q_{\text{TL}}\big(\bar{\tilde{f}}\big)\,C_{\beta\epsilon}\bar{\tilde{f}}_{\epsilon}\\
 & =A\big(\bar{\tilde{f}}\big)_{\alpha\beta}-\frac{2}{N}Q_{\text{TL}}^{2}\big(\bar{\tilde{f}}\big)\,\left[C\bar{\tilde{f}}\bar{\tilde{f}}^{\T}C\right]_{\alpha\beta},\label{eq:Hessian_action_tl}
\end{align}
where we use the shorthand $A\big(\bar{\tilde{f}}\big)=\kappa_{0}\mathbb{I}+Q_{\text{TL}}\big(\bar{\tilde{f}}\big)\,C$.
Using the relation between $\tilde{f}$ and the mean deviations from
\eqref{eq:relation_Delta_tilde_f_stats} $\langle\Delta_{\alpha}\rangle=i\kappa_{0}\bar{\tilde{f}}_{\alpha}$,
we may rewrite this expression as
\begin{equation}
\frac{\partial^{2}(-\mathcal{S}(\bar{\tilde{f}},j_{\ast}))}{\partial\bar{\tilde{f}}_{\alpha}\partial\bar{\tilde{f}}_{\beta}}\bigg\vert_{j^{*}=0}=A\big(\bar{\tilde{f}}\big)_{\alpha\beta}+\frac{2}{N}Q_{\text{TL}}^{2}\big(\bar{\tilde{f}}\big)\,\kappa_{0}^{-2}\left[C_{\mathcal{DD}}^{(xx)}\langle\Delta\rangle\langle\Delta\rangle^{\top}C_{\mathcal{DD}}^{(xx)}\right]_{\alpha\beta},
\end{equation}
which yields the expression from the main text \eqref{eq:cov_adaptive}
for the covariance of the network output on training data with $\kappa=\kappa_{0}N^{1-\gamma}$
and hence
\begin{equation}
\langle\langle ff^{\top}\rangle\rangle=\kappa\mathbb{I}-\kappa^{2}\Bigg(A\big(\bar{\tilde{f}}\big)+\frac{2Q_{\text{TL}}^{2}}{N\kappa_{0}^{2}}F\Bigg)\label{eq:cov_outputs_final}
\end{equation}
with $F=C_{\mathcal{DD}}^{(xx)}\langle\Delta\rangle\langle\Delta\rangle^{\T}C_{\mathcal{DD}}^{(xx)}$.

\subsubsection{Non-linear activation function}

Here we consider a point-symmetric non-linear activation function,
so that the mean $m=0$; the extension to the case with $m\neq0$
is straightforward. Due to the similarity of the expressions \eqref{eq:W_Gauss}
and \eqref{eq:W_nonlin_final}, the final expressions here have the
same structure as in the case of the linear activation function, Eqs.
\eqref{eq:Delta_tree_level} - \eqref{eq:f_tilde_tree_level}, but
with the replacement $C^{(xx)}\to C^{(\phi\phi)}$ throughout. Because
these two cases throughout lead to identical expressions except for
this replacement, in the following we denote $C^{(xx)}$ or $C^{(\phi\phi)}$
simply by $C$.

\subsubsection{Non-linear activation function - beyond cumulant expansion \label{app:VGA}}

In this section we consider a more fine-grained approach compared
to the cumulant expansion in the previous section. Following results
from previous sections, the self-consistency equation for $\bar{\tilde{f}}$
in a non-linear network is given by
\begin{equation}
\bar{\tilde{f}}=-i\left(K+\kappa_{0}/N\mathbb{I}\right)^{-1}y
\end{equation}
with 
\begin{equation}
K=g_{w}/N\frac{\int d^{P}h\,\phi\left(h\right)\phi\left(h\right)^{\T}\exp\left(-\frac{1}{2}h^{T}C^{\left(xx\right)}h-\frac{g_{w}}{2N^{2}}\bar{\tilde{f}}^{\T}\phi\left(h\right)\phi\left(h\right)^{T}\bar{\tilde{f}}\right)}{\int d^{P}h\exp\left(-\frac{1}{2}h^{\T}C^{\left(xx\right)}h-\frac{g_{w}}{2N^{2}}\bar{\tilde{f}}^{\T}\phi\left(h\right)\phi\left(h\right)^{T}\bar{\tilde{f}}\right)}.\label{eq:kernel_definition}
\end{equation}
In the case of linear networks, the integrals in \prettyref{eq:kernel_definition}
are tractable, and are replaced in previous sections by their explicit
solution, as can be seen in \eqref{eq:cum_gen_linear}. In the case
of non-linear networks, the value of $K$ must be approximated, either
via a cumulant-expansion as in the previous section, via a variational
Gaussian approximation (VGA) as in \cite{seroussi23_908}, or via
higher-order distributional approximations such as the variational
Gaussian mixture approximation as in \cite{Rubin24_iclr}, as well
as potentially richer approximations. Here we consider the VGA, but
we emphasize that our framework is in no way limited to a Gaussian
assumption.

\begin{figure*}[t]
\includegraphics[width=1\textwidth]{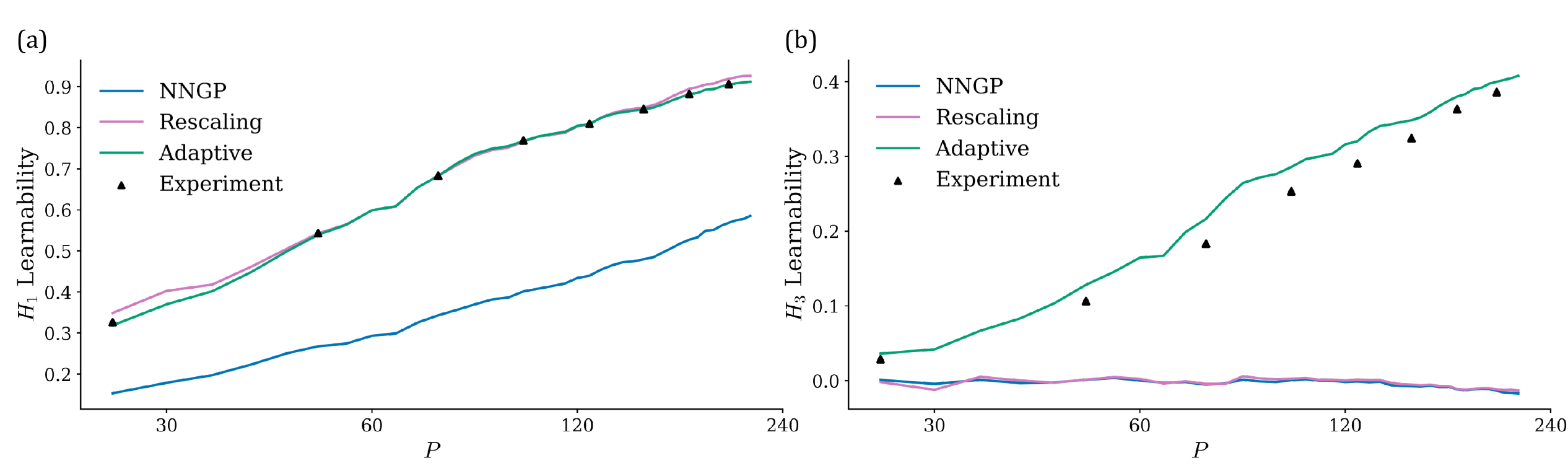}\vspace{0.2in}

\centering{}\caption{$g$-learnability of target components in two-layer erf network, trained
on the target given in \prettyref{eq:teacher}, where $H_{i}$ learnability
is defined according to \prettyref{eq:learnability}, with $g(X)=H_{i}(Xw_{*})$.
As can be seen in panel (b), the adaptive approach as derived using
VGA predicts that the network will begin to learn higher-order components
of $y$ at $P\sim\mathcal{O}(D)$, due to the manifestation of directionally
dependent feature learning. On the other hand, kernel methods such
as the NNGP and the rescaling approach predict that the network output
will be linear, and the cubic component of the target would require
$P\sim(D^{3})$ training samples. Parameters: $P_{\text{test}}=4.000$,
$N=1.000$, $D=30$, $\kappa_{0}=2$, $g_{v}=g_{w}=1,$ $\gamma=1,\epsilon=-0.1$.\label{fig:complexity_change_erf}}
\vspace{0.2in}
\end{figure*}

The VGA defines a matrix $\Sigma$ which is chosen such that it minimizes
the KL divergence between the distribution of $h$ as it appears in
\prettyref{eq:kernel_definition} and a Gaussian distribution with
covariance $\Sigma$. The matrix $\Sigma$ is then determined by the
following equation
\begin{equation}
\left[\Sigma^{-1}\right]_{ij}=\left[C^{\left(xx\right)}\right]_{ij}-y^{\T}\left(K_{\Sigma}+\kappa_{0}/N\mathbb{I}\right)^{-1}\frac{\partial K_{\Sigma}}{\partial\Sigma_{ij}}\left(K_{\Sigma}+\kappa_{0}/N\mathbb{I}\right)^{-1}y\label{eq:Sigma_vga}
\end{equation}
where $K_{\Sigma}:=g_{w}/N\left\langle \phi\left(h\right)\phi\left(h\right)^{\T}\right\rangle _{h\sim\mathcal{N}\left(0,\Sigma\right)}$
(which is tractable for activations such as Erf, ReLU). Thus the equation
for $\bar{\tilde{f}}$ can be written as
\begin{equation}
\bar{\tilde{f}}_{\text{VGA}}=-i\left(K_{\Sigma}+\kappa_{0}/N\mathbb{I}\right)^{-1}y.
\end{equation}
We note that for a linear network, the VGA approximation is exact,
as the distribution of $h$ is indeed Gaussian. The application of
the VGA approximation allows for structural changes to the kernel
which do not emerge from a simple cumulant expansion. The structural
encoding of certain directions in the kernel could allow the network
to learn complex functions of these directions with significantly
less data points than would be naively expected. Thus, even a small
non-linearity could lead to fundamentally different predictions by
the adaptive approach compared to rescaling approaches. To demonstrate
this, we consider a teacher-student setting, where the teacher is
given by
\begin{equation}
y(x)=H_{1}(w_{*}^{\T}x)+\epsilon H_{3}(w_{*}^{\T}x),\label{eq:teacher}
\end{equation}
with $H_{1,3}$ being the first- and third-order Hermite polynomials,
and $x\sim\mathcal{N}(0,\mathbb{I})$. We quantify the ability of
a network to learn the different target components (linear or cubic)
with a parameter we call \emph{learnability}, defined so that $g$-learnability
is given by 
\begin{equation}
g\text{-learnability}:=\frac{f\left(X_{\text{test}}\right)^{\T}g\left(X_{\text{test}}\right)}{y\left(X_{\text{test}}\right)^{\T}g\left(X_{\text{test}}\right)}\,,\label{eq:learnability}
\end{equation}
where $f,\,y,\,g$ are applied row-wise, and $g$ is any function.
Having a learnability = 1 for a given component implies that the network
has successfully learned this component. In \prettyref{fig:complexity_change_erf}
we show the theoretical predictions as well as experimental observations
for the learnability of both target components ($H_{1,3}(w_{*}^{\T}x)$).
The experimentally measured learnability of the non-linear component
is in good agreement with the adaptive predictions (obtained by VGA),
which significantly outperforms the predictions of the kernel approaches
(both rescaling and NNGP).

\subsection{One-Loop corrections in standard scaling\label{app:one_loop_app}}

While in mean-field scaling ($\gamma=2$) the cumulant-generating
function has a scaling form and the network outputs $f$ concentrate,
we need to account for their fluctuations in standard scaling ($\gamma=1$).
In the following, we thus set $\gamma=1$. To leading order, also
called one-loop approximation, we have
\begin{equation}
\Gamma_{\text{1-Loop}}(\bar{\tilde{f}},j_{\ast}\vert y)=-\mathcal{S}(\bar{\tilde{f}},j_{\ast})-\frac{1}{2}\log\det(-\mathcal{S}^{(2)}).
\end{equation}
The self-consistency equation for $\bar{\tilde{f}}$ then becomes
\begin{equation}
\frac{\partial\Gamma_{\text{1-Loop}}(\bar{\tilde{f}},j_{\ast}\vert y)}{\partial\bar{\tilde{f}}_{\delta}}\bigg\vert_{j^{*}=0}=-\frac{\partial\mathcal{S}(\bar{\tilde{f}},j_{\ast})}{\partial\bar{\tilde{f}}_{\delta}}\bigg\vert_{j^{*}=0}-\frac{1}{2}\sum_{\alpha\beta}(-\mathcal{S}^{(2)})_{\beta\alpha}^{-1}\frac{\partial^{3}(-\mathcal{S}(\bar{\tilde{f}},j_{\ast}))}{\partial\bar{\tilde{f}}_{\alpha}\partial\bar{\tilde{f}}_{\beta}\partial\bar{\tilde{f}}_{\delta}}\bigg\vert_{j^{*}=0}\overset{!}{=}0.
\end{equation}
Given the form \eqref{eq:W_Gauss} of the cumulant-generating function
$W$, taking the derivative of $\mathcal{S}$ with respect to $\bar{\tilde{f}}$
in \eqref{eq:eq_state_tilde_f} yields a different expression than
in the previous section
\begin{align}
-\frac{\partial\mathcal{S}(\bar{\tilde{f}},j_{\ast})}{\partial\bar{\tilde{f}}_{\rho}}\bigg\vert_{j^{*}=0} & =iy_{\rho}+\kappa_{0}\bar{\tilde{f}}_{\rho}+Q_{\text{TL}}\big(\bar{\tilde{f}}\big)\,C_{\rho\delta}\overline{\tilde{f}}_{\delta},\label{eq:stationary_tilde_f}\\
Q_{\text{TL}}\big(\bar{\tilde{f}}\big) & =\frac{g_{w}}{1+\frac{g_{w}}{N^{\gamma}}\bar{\tilde{f}}^{\T}C\bar{\tilde{f}}},
\end{align}
where now $Q_{\text{TL}}$ is a scalar. Note that in this form the
tree-level equation for $\bar{\tilde{f}}$ \eqref{eq:f_tilde_tree_level}
can be written as
\begin{equation}
\bar{\tilde{f}}=-i\left(\kappa_{0}N^{1-\gamma}\mathbb{I}+Q_{\text{TL}}\big(\bar{\tilde{f}}\big)\,C\right)^{-1}y,
\end{equation}
thereby obtaining an expression in which the input kernel $C$ is
only rescaled by a scalar, which we call a kernel rescaling expression.
For the second and third derivatives, we obtain
\begin{align}
\frac{\partial^{2}(-\mathcal{S}(\bar{\tilde{f}},j_{\ast}))}{\partial\bar{\tilde{f}}_{\alpha}\partial\bar{\tilde{f}}_{\beta}}\bigg\vert_{j^{*}=0} & =\kappa_{0}\delta_{\alpha\beta}+Q_{\text{TL}}\big(\bar{\tilde{f}}\big)\,C_{\alpha\beta}-\frac{2}{N}Q_{\text{TL}}\big(\bar{\tilde{f}}\big)\,C_{\alpha\delta}\bar{\tilde{f}}_{\delta}\,Q_{\text{TL}}\big(\bar{\tilde{f}}\big)\,C_{\beta\epsilon}\bar{\tilde{f}}_{\epsilon}\\
 & =A\big(\bar{\tilde{f}}\big)_{\alpha\beta}-\frac{2}{N}Q_{\text{TL}}^{2}\big(\bar{\tilde{f}}\big)\,\left[C\bar{\tilde{f}}\bar{\tilde{f}}^{\T}C\right]_{\alpha\beta},\label{eq:Hessian_action}\\
\frac{\partial^{3}(\mathcal{-S}(\bar{\tilde{f}},j_{\ast}))}{\partial\bar{\tilde{f}}_{\alpha}\partial\bar{\tilde{f}}_{\beta}\partial\bar{\tilde{f}}_{\text{\ensuremath{\delta}}}}\bigg\vert_{j^{*}=0} & =-\frac{2}{N}Q_{\text{TL}}^{2}\big(\bar{\tilde{f}}\big)\,\left[C_{\alpha\beta}C_{\delta\epsilon}\bar{\tilde{f}}_{\epsilon}+C_{\alpha\delta}C_{\beta\epsilon}\bar{\tilde{f}}_{\epsilon}+C_{\beta\delta}C_{\alpha\epsilon}\bar{\tilde{f}}_{\epsilon}\right]\\
 & \phantom{=}+\frac{8}{N^{2}}Q_{\text{TL}}^{3}\big(\bar{\tilde{f}}\big)\,C_{\alpha\alpha^{\prime}}\bar{\tilde{f}}_{\alpha^{\prime}}\,C_{\beta\beta^{\prime}}\bar{\tilde{f}}_{\beta^{\prime}}\,C_{\delta\delta^{\prime}}\bar{\tilde{f}}_{\delta^{\prime}}\\
 & =-\frac{2}{N}Q_{\text{TL}}^{2}\big(\bar{\tilde{f}}\big)\,\left(C_{\alpha\beta}\left[C\bar{\tilde{f}}\right]_{\delta}+C_{\alpha\delta}\left[C\bar{\tilde{f}}\right]_{\beta}+C_{\beta\delta}\left[C\bar{\tilde{f}}\right]_{\alpha}\right)\\
 & \phantom{=}+\frac{8}{N^{2}}Q_{\text{TL}}^{3}\big(\bar{\tilde{f}}\big)\,\left[C\bar{\tilde{f}}\right]_{\alpha}\left[C\bar{\tilde{f}}\right]_{\beta}\left[C\bar{\tilde{f}}\right]_{\delta}.
\end{align}
Here, we use the shorthand $A\big(\bar{\tilde{f}}\big)=\kappa_{0}\mathbb{I}+Q_{\text{TL}}\big(\bar{\tilde{f}}\big)\,C$.
Overall, we obtain
\begin{align}
\bar{\tilde{f}}_{\delta} & =\Big[A\big(\bar{\tilde{f}}\big)^{-1}\Big]_{\delta\epsilon}\Bigg[-iy_{\epsilon}+\frac{1}{2}\sum_{\alpha\beta}(-\mathcal{S}^{(2)})_{\beta\alpha}^{-1}\frac{\partial^{3}(-\mathcal{S}(\bar{\tilde{f}},j_{\ast}))}{\partial\bar{\tilde{f}}_{\alpha}\partial\bar{\tilde{f}}_{\beta}\partial\bar{\tilde{f}}_{\epsilon}}\bigg\vert_{j^{*}=0}\Bigg].
\end{align}
Similarly to the previous section, the training discrepancies are
given by
\begin{equation}
\langle\Delta_{\alpha}\rangle=i\kappa_{0}\bar{\tilde{f}}_{\alpha}.
\end{equation}
For the test predictor, we have
\begin{align}
\langle f^{*}\rangle_{\mathrm{\text{1-Loop}}} & =-\frac{\partial\Gamma_{\text{1-Loop}}(\bar{\tilde{f}},j_{\ast}\vert y)}{\partial j^{*}}\bigg\vert_{j^{*}=0}\\
 & =-\frac{\partial W(\bar{\tilde{f}},j_{\ast})}{\partial j^{*}}\bigg\vert_{j^{*}=0}-\frac{1}{2}\sum_{\alpha\beta}(-\mathcal{S}^{(2)})_{\beta\alpha}^{-1}\frac{\partial^{3}(-\mathcal{S}(\bar{\tilde{f}},j_{\ast}))}{\partial\bar{\tilde{f}}_{\alpha}\partial\bar{\tilde{f}}_{\beta}\partial j_{\ast}}\bigg\vert_{j^{*}=0}\\
 & =Q_{\text{TL}}\big(\bar{\tilde{f}}\big)\,C_{*\alpha}\bar{\tilde{f}}_{\alpha}-\frac{1}{2}\sum_{\alpha\beta}(-\mathcal{S}^{(2)})_{\beta\alpha}^{-1}\frac{\partial^{3}(-\mathcal{S}(\bar{\tilde{f}},j_{\ast}))}{\partial\bar{\tilde{f}}_{\alpha}\partial\bar{\tilde{f}}_{\beta}\partial j_{\ast}}\bigg\vert_{j^{*}=0}.
\end{align}
The appearing derivatives of the action are structurally similar but
we replace the training point $x_{\gamma}$ by the test point $x_{\ast}$,
yielding
\begin{align}
\frac{\partial^{3}(-\mathcal{S}(\bar{\tilde{f}},j_{\ast}))}{\partial\bar{\tilde{f}}_{\alpha}\partial\bar{\tilde{f}}_{\beta}\partial j_{\ast}}\bigg\vert_{j^{*}=0} & =-\frac{2}{N}Q_{\text{TL}}^{2}\big(\bar{\tilde{f}}\big)\,\Big(C_{\alpha\beta}\left[C\bar{\tilde{f}}\right]_{\ast}+C_{\alpha\ast}\left[C\bar{\tilde{f}}\right]_{\beta}+C_{\beta\ast}\left[C\bar{\tilde{f}}\right]_{\alpha}\Big)\\
 & \phantom{=}+\frac{8}{N^{2}}Q_{\text{TL}}^{3}\big(\bar{\tilde{f}}\big)\,\left[C\bar{\tilde{f}}\right]_{\alpha}\left[C\bar{\tilde{f}}\right]_{\beta}\left[C\bar{\tilde{f}}\right]_{\ast}.
\end{align}
When solving these equations, we backtransform to the imaginary variables
$\bar{\tilde{f}}\mapsto i\bar{\tilde{f}}$, which changes multiple
signs and absorbs the appearing imaginary units.

\subsection{Kernel rescaling approach\label{app:kernel_scaling_app}}

We here derive the results by \citet{Li21_031059} and \citet{Ariosto2022}
in our multi-scale adaptive theory: Using that $h_{\alpha j}\sim\N(0,C^{(xx)})$
i.i.d. over the neuron index $j$, we can rewrite the cumulant-generating
function $W$ \eqref{eq:cum_gen_output} for the case of a linear
activation function $\phi(h)=h$ conditioned on readout weights $w$
as
\begin{align}
W(\tilde{f}_{\mathcal{D}}\vert w) & =\ln\left\langle \exp(-\tilde{f_{\alpha}}w_{j}h_{\alpha j})\right\rangle _{h_{\alpha j}}=\frac{1}{2}\tilde{f_{\alpha}}C_{\alpha\beta}^{(xx)}\tilde{f}_{\beta}\,\|w\|^{2},
\end{align}
where we drop the test point here to keep notation concise. The result
for the test point will follow naturally later. Likewise, performing
a cumulant expansion up to second order in $\phi$ for a point-symmetric
activation function as in \eqref{eq:W_non_lin}, we obtain 
\begin{align}
W(\tilde{f}_{\mathcal{D}}\vert w) & =\ln\left\langle \exp(\tilde{f_{\alpha}}w_{j}\phi(h_{\alpha j})\right\rangle _{h_{\alpha j}}=\frac{1}{2}\tilde{f_{\alpha}}C_{\alpha\beta}^{(\phi\phi)}\tilde{f}_{\beta}\,\|w\|^{2},
\end{align}
where $C^{(\phi\phi)}$ is defined as in \eqref{eq:cum_phi}. As in
the adaptive approach, we here again write $C$ for short to refer
to $C^{(xx)}$ in the case of linear activation function and to $C^{(\phi\phi)}$
in the case of the non-linear point symmetric activation.

We observe that the readout weights only appear in the form of the
squared norm $\|w\|^{2}$. The distribution of the network output
is hence
\begin{align}
p(y,f|C) & =\N(y|f,\kappa_{0})\,\int\mathrm{d}\tilde{f}_{\mathcal{D}}\,\big\langle\exp\big(-i\tilde{f}_{\alpha}f_{\alpha}-\frac{1}{2}\tilde{f_{\alpha}}C_{\alpha\beta}\tilde{f}_{\beta}\,\|w\|^{2}\big)\big\rangle_{w_{i}\stackrel{\text{i.i.d. }}{\sim}\N(0,\frac{g_{w}}{N})}.\label{eq:Appendix_NetworkPosterior_Sompolinsky}
\end{align}
Since both, the prior measure of the weights $w\sim\N(w|0,g_{w}N^{-1})\propto\exp(N\,\|w\|^{2}/2g_{w})$
and the explicit appearance of $w$, is only in the form of $\|w\|^{2}$,
we may introduce this quantity as an auxiliary variable, which we
name $Q:=\|w\|^{2}=\sum_{i=1}^{N}w_{i}^{2}$ and which corresponds
to the Euclidean norm of the readout weight vector $w$. Note that,
given $\|w\|^{2}$, the integral over $\tilde{f}_{\mathcal{D}}$ simply
yields $f|_{\|w\|^{2}}\sim\N(0,\|w\|^{2}C)$, so
\begin{align}
p(y,f|C) & =\N(y|f,\kappa_{0})\,\int dQ\,\N(f|0,Q\,C)\,p(Q).\label{eq:p_z_y_Q}
\end{align}
Here the distribution of the squared norm is 
\begin{align}
p(Q) & =\big\langle\delta[-Q+\|w\|^{2}]\big\rangle_{w_{i}\stackrel{\text{i.i.d.}}{\sim}\N(0,\frac{g_{w}}{N})}\label{eq:p_Q}\\
 & =\int_{-i\infty}^{i\infty}\frac{d\tilde{Q}}{2\pi i}\,\big\langle\exp\big(\tilde{Q}\big[-Q+\|w\|^{2}\big]\big)\big\rangle_{w_{i}\stackrel{\text{i.i.d.}}{\sim}\N(0,\frac{g_{w}}{N})}\\
 & =\int_{-i\infty}^{i\infty}\frac{d\tilde{Q}}{2\pi i}\,\exp\big(-\tilde{Q}\,Q+W(\tilde{Q})\big),
\end{align}
where $W(\tilde{Q})=\ln\,\langle\exp\big(\tQ\,\|w\|^{2})\rangle_{w_{i}\stackrel{\text{i.i.d.}}{\sim}\N(0,\frac{g_{w}}{N})}$
is the cumulant-generating function of $Q$ . Using that the $w_{j}$
are i.i.d, we get
\begin{align}
W(\tQ) & =N\,\ln\,\big\langle\exp\big(\tQ\|w\|^{2}\big)\big\rangle_{w\sim\N(0,g_{w}/N)}\\
 & =-\frac{N}{2}\,\ln\big[1-\frac{2g_{w}}{N}\tilde{Q}\big],
\end{align}
where we performed the one-dimensional Gaussian integral over $w$.
Up to here, all steps are exact.

\prettyref{eq:p_z_y_Q} shows that the auxiliary variable $Q$ being
a scalar may only carry fluctuations of the overall scaling of the
kernel and hence all descriptions and approximations in terms of $Q$
can only change the scale of the kernel, which is consistent with
the results in \cite{Li21_031059,Pacelli23_1497}.

\subsubsection{Approximation of network prior for wide networks}

One expects that $Q$ concentrates for large $N$ according to the
central limit theorem since $Q=\|w\|^{2}=\sum_{i=1}^{N}w_{i}^{2}$
with i.i.d. $w_{i}\sim\N(0,g_{w}/N)$. The cumulant-generating function
$W$ can be written as a scaling form $\lambda_{N}(k):=N^{-1}\,W(N\,k)=-\frac{1}{2}\,\ln\big[1-2g_{w}k\big]$
and its limit $N\to\infty$ then exists trivially, so that we may
approximate $p(Q)$ with the G\"artner-Ellis theorem \cite{Touchette09}
as
\begin{align}
\ln\,p(Q) & \simeq\sup_{\tQ}\,-Q\tQ+W(\tQ)\label{eq:GammaQ}\\
 & =-\frac{N}{2g_{w}}\,\big[1-\frac{g_{w}}{Q}\big]\,Q-\frac{N}{2}\,\ln\big[\frac{g_{w}}{Q}\big]\\
 & =-\frac{N}{2}\,\big[\frac{Q}{g_{w}}-1-\ln\,\frac{Q}{g_{w}}\big]=:-\Gamma(Q).
\end{align}
Intuitively, by the scaled cumulant-generating function of the form
$N\,W(\tilde{Q}/N)=-\frac{N}{2}\,\ln\big[1-2g_{w}\,\frac{\tilde{Q}}{N}\big]$
the mean of is of order $\langle Q\rangle=\mathcal{O}(1)$ and all
higher-order cumulants of $Q$ are being suppressed by at least $\mathcal{O}(N^{-1})$.
So on exponential scales, one may parametrize the probability by the
mean, namely one obtains the distribution of $Q$ from the rate function
as $e^{-\Gamma(Q)}$. To obtain \prettyref{eq:GammaQ}, the supremum
condition has been used $0\stackrel{!}{=}-Q+g_{w}\,\big[1-2g_{w}\frac{\tilde{Q}}{N}\big]^{-1}$,
solved for $1-\frac{2g_{w}}{N}\,\tQ=\frac{g_{w}}{Q}$ and $\tilde{Q}=\frac{N}{2g_{w}}\,\big[1-\frac{g_{w}}{Q}\big]$
and inserted into the first line of \prettyref{eq:GammaQ} to obtain
the second line. The rate function, being the Legendre transform of
$W$, obeys the equation of state
\begin{align}
\frac{d}{dQ}\Gamma(Q) & =\tQ=\frac{N}{2g_{w}}\,\big[1-\frac{g_{w}}{Q}\big].\label{eq:eos_Gamma}
\end{align}
So the final expression for the joint probability of $y$ and $f$,
the network prior, is
\begin{align}
p(y,f|C) & \simeq\N(y|f,\kappa_{0})\,\int dQ\,\N(f|0,Q\,C)\,e^{-\Gamma(Q)},\label{eq:p_z_y_pre_saddle-1}\\
 & =\int dQ\,e^{S(Q|f,y)},\nonumber 
\end{align}
where the action $S(Q|f,y)$ is
\begin{align}
S(Q|f,y) & =-\frac{\|y-f\|^{2}}{2\kappa_{0}}-\frac{P}{2}\ln\kappa_{0}\label{eq:action_Q-2-1}\\
 & \phantom{=}-\frac{1}{2}f^{\T}\,\big(QC\big)^{-1}\,f-\frac{1}{2}\ln\det\big(QC\big)-\Gamma(Q)+\mathrm{const}.\nonumber 
\end{align}

\subsubsection{Maximum a posteriori estimate for $Q$}

To obtain the posterior distribution for $Q$ we marginalize \eqref{eq:p_z_y_pre_saddle-1}
over the network outputs $f$, which yields
\begin{align}
p(y|C) & \equiv\int df\,p(y,f|C)\label{eq:p_z_Q-1}\\
 & =\int dQ\,\exp\big(S(Q|y)\big),\nonumber 
\end{align}
which yields the action
\begin{align}
S(Q|y) & =-\frac{1}{2}y^{\T}\,\big(QC+\kappa_{0}\I\big)^{-1}\,y-\frac{1}{2}\ln\det\big(QC+\kappa_{0}\I\big)-\Gamma(Q),\label{eq:S_Q_z-1}
\end{align}
and which reproduces Eq. A11 in \citet{Li21_031059} after inserting
the rate function \eqref{eq:GammaQ} and using $C=C^{(xx)}$ for the
linear network. It likewise reproduces Eq. (33) in \cite{Ariosto2022}
when inserting $C=C^{(\phi\phi)}$ for the non-linear activation function.

When computing the maximum a posteriori value $Q_{\text{LS}}$, it
only depends on the numerator of
\begin{align}
p(Q|y) & =\frac{p(y|Q)\,p(Q)}{p(y)},
\end{align}
since the form of \eqref{eq:p_z_Q-1} is $p(y)=\int dQ\,p(y|Q)\,p(Q)$.
Thus, computing the $Q$-integral in saddle point approximation comprises
to the maximum a posteriori (MAP) $Q_{\text{LS}}$ as $\ln\,p(y|Q)\,p(Q)=S(Q|y)+\const$
has the same stationary point as $p(y|Q)\,p(Q)$.

The length $Q=\|w\|^{2}$ in their theory is obtained by the maximum
of \eqref{eq:S_Q_z-1}, which is given by
\begin{align}
0 & \overset{!}{=}\frac{\partial S}{\partial Q}=\frac{1}{2}y^{\T}\big(QC+\kappa_{0}\I\big)^{-1}C\,\big(QC+\kappa_{0}\I\big)^{-1}y\label{eq:stat_Q-1}\\
 & \qquad\qquad\quad-\tr\,C\big(QC+\kappa_{0}\I\big)^{-1}-\frac{N}{2}\big(\frac{1}{g_{w}}-\frac{1}{Q}\big).\nonumber 
\end{align}
This self-consistency equation yields the tree-level approximation
for $\|w\|^{2}=Q_{\text{LS}}$.

\subsubsection{Predictor statistics\label{sec:Predictor-statistics}}

To obtain predictions beyond the length of the readout $\|w\|$, we
start from \ref{eq:p_z_Q-1}. We obtain statistics of the training
discrepancies $\Delta=y_{\alpha}-f_{\alpha}$ from
\begin{align}
\frac{\partial}{\partial y_{\alpha}}\,\ln p(y|C)\stackrel{\text{MAP }Q}{\simeq} & \frac{d}{dy_{\alpha}}\,\sup_{Q}S(Q|y)\\
= & \frac{\partial}{\partial y_{\alpha}}\,S(Q_{\text{LS}}|y)+\underbrace{\frac{\partial S}{\partial Q}}_{=0}\,\frac{\partial Q}{\partial y_{\alpha}}\big|_{Q=Q^{\ast}},
\end{align}
where the derivative by $Q$ vanishes because $Q_{\text{LS}}$ has
been determined by the supremum condition as the stationary point
of the action. The partial derivative by $y_{\alpha}$ only acts on
$-y^{\T}\,\big(QC+\kappa_{0}\I\big)^{-1}\,y/2$ in the expression
for \eqref{eq:S_Q_z-1}
\begin{align}
\langle\Delta_{\alpha}\rangle & =\kappa_{0}\,\big(Q_{\text{LS}}C+\kappa_{0}\I\big)^{-1}\,y.\label{eq:mean_discrepancies}
\end{align}
In consequence, the test predictor is identical to the NNGP predictor
with a different regularizer $\kappa_{0}/Q_{\text{LS}}$
\begin{align}
\langle f_{\ast}\rangle_{\text{LS}} & =\big[C_{\ast\alpha}\big]\,\big(C+\kappa_{0}/Q_{\text{LS}}\I\big)_{\alpha\beta}^{-1}\,y_{\beta}.
\end{align}

To compute the variance of the predictor, we generalize\eqref{eq:Appendix_NetworkPosterior_Sompolinsky}
such that instead of the variance $\kappa_{0}\I$ in $\N(y|f,\kappa_{0}\I)$,
we insert a general covariance matrix $K$ into the Gaussian measure
$\N(y|f,K)$ and perform an integration over $f$
\begin{align}
p(y|K,C^{(xx)}) & =\int df\,\N(y|f,K)\,\big\langle\,\prod_{\alpha=1}^{P}\delta\,\big[f_{\alpha}-\sum_{i=1}^{N}w_{i}\,\phi(h_{\alpha i})\big]\big\rangle_{w_{i}\stackrel{\text{i.i.d.}}{\sim}\N(0,\frac{g_{w}}{N}),\quad h_{\alpha i}\stackrel{\text{i.i.d. over }i}{\sim}\N(0,C^{(xx)})}.\label{eq:p_y_z_given_X-1}
\end{align}
The presence of the general matrix $K$ allows us to measure the statistics
of the discrepancies $\Delta_{\alpha}=y_{\alpha}-z_{\alpha}$, because
writing the Gaussian $\N(y|f,K)\propto\exp\big(-\frac{1}{2}(y-f)^{\T}K^{-1}(y-f)+\frac{1}{2}\,\ln\,\det\,(K^{-1})\big)$
explicitly we observe that derivatives by $\big[K^{-1}\big]_{\alpha\beta}$
yield
\begin{align}
\frac{\partial}{\partial[K]_{\alpha\beta}^{-1}}\ln p(y|K,C)\Big|_{K=\kappa_{0}\I} & =-\frac{1}{2}\langle(y-f)_{\alpha}(y-f)_{\beta}\rangle+\frac{1}{2}\,\kappa_{0}\,\delta_{\alpha\beta}.\label{eq:gen_disc}
\end{align}
With the same manipulations that led to \eqref{eq:p_z_Q-1} one then
has
\begin{align}
p(y|K,C) & \equiv\int dQ\,\exp\big(S(Q|K,z)\big),
\end{align}
where the action, corresponding to \eqref{eq:S_Q_z-1}, is
\begin{align}
S(Q|y,K) & =-\frac{1}{2}y^{\T}\,\big(C+K\big)^{-1}\,y-\frac{1}{2}\ln\det\big(C+K\big)-\frac{N}{2}\,\big(\frac{Q}{g_{w}}-\ln\,Q\Big)\big|_{C=QC}.
\end{align}
So in the approximation replacing $Q$ by its MAP $Q_{\text{LS}}$
we get
\begin{align}
\frac{\partial}{\partial[K]_{\alpha\beta}^{-1}}\ln p(y|K,C)\Big|_{K=\kappa_{0}\I} & =\frac{d}{d[K]_{\alpha\beta}^{-1}}\,\sup_{Q}S(Q|y,K)\Big|_{K=\kappa_{0}\I}\\
 & =\frac{\partial}{\partial[K]_{\alpha\beta}^{-1}}S(Q_{\text{LS}}|y,K)\Big|_{K=\kappa_{0}\I},
\end{align}
where the inner derivative by $\partial S/\partial Q$ drops out due
to stationarity at $Q_{\text{LS}}$, which is given by the solution
of \eqref{eq:stat_Q-1}. The latter partial derivative evaluates to
\begin{align}
\frac{\partial}{\partial[K]_{\alpha\beta}^{-1}}S(Q|y,K)\Big|_{K=\kappa_{0}\I} & =\Big[-\frac{1}{2}K\,\big[c+K\big]^{-1}yy^{\T}\,\big[c+K\big]^{-1}\,K+\frac{1}{2}\,K\,(c+K)^{-1}\,K\Big]_{\alpha\beta}\Big|_{K=\kappa_{0}\I,\,c=Q_{\text{LS}}\,C}\\
 & =\kappa_{0}^{2}\,\Big[-\frac{1}{2}\,\big[c+\kappa_{0}\I\big]^{-1}yy^{\T}\,\big[c+\kappa_{0}\I\big]^{-1}+\frac{1}{2}\,(c+\kappa_{0}\I)^{-1}\Big]_{\alpha\beta}\Big|_{c=Q_{\text{LS}}\,C},
\end{align}
where we used that $\partial K_{\gamma\delta}/\partial[K]_{\alpha\beta}^{-1}=-K_{\gamma\alpha}\,K_{\beta\delta}$,
which follows by symmetry from $\partial K_{\gamma\delta}^{-1}/\partial K_{\alpha\beta}=-K_{\gamma\alpha}^{-1}\,K_{\beta\delta}^{-1}$.

So the second moment of the discrepancies with \eqref{eq:gen_disc}
is
\begin{align}
\langle\Delta_{\alpha}\Delta_{\beta}\rangle & =\kappa_{0}\delta_{\alpha\beta}+\kappa_{0}^{2}\,\Big[\,\big[c+\kappa_{0}\I\big]^{-1}yy^{\T}\,\big[c+\kappa_{0}\I\big]^{-1}-(c+\kappa_{0}\I)^{-1}\Big]_{\alpha\beta}\Big|_{c=Q_{\text{LS}}\,C}\nonumber \\
 & =\langle\Delta_{\alpha}\rangle\langle\Delta_{\beta}\rangle+\kappa_{0}\delta_{\alpha\beta}-\kappa_{0}^{2}\,(c+\kappa_{0}\I)_{\alpha\beta}^{-1}\Big|_{c=Q_{\text{LS}}\,C},\label{eq:second_moment_Delta}
\end{align}
where we used \eqref{eq:mean_discrepancies} in the last step. Because
$\Delta=y-f$ and the target label $y$ do not fluctuate, the latter
two terms in \eqref{eq:second_moment_Delta} are the variance
\begin{eqnarray}
\llangle\Delta_{\alpha},\Delta_{\beta}\rrangle & = & \llangle f_{\alpha},f_{\beta}\rrangle\label{eq:cov_rescaling}\\
 & = & \kappa_{0}\delta_{\alpha\beta}-\kappa_{0}^{2}\,(c+\kappa_{0}\I)_{\alpha\beta}^{-1}\,\Big|_{c=Q_{\mathrm{LS}}\,C}\\
 & = & c-c\,[c+\kappa_{0}\I]^{-1}\,c\,\Big|_{c=Q_{\mathrm{LS}}\,C},
\end{eqnarray}
which is the usual expression for the variance of the NNGP predictor
of a Gaussian process with the kernel $c=Q_{\text{LS}}\,C$.

\subsection{Connecting kernel rescaling and adaptive approach\label{app:connection_adpative_kernel_app}}

While the kernel rescaling approach holds in the proportional limit
${N\propto P\rightarrow\infty}$, the one-loop approximation holds
also for large but finite $P,N\gg$1. As we have seen in \prettyref{fig:scatter_standard_scaling_nngp}
in the main text, they yield almost identical results in certain settings.
By considering the proportional limit, we may connect these two approaches:
some correction terms vanish in this limit, leaving only a scalar
term.

To this end, we look at the scaling of each correction term with both
$P$ and $N$. We have
\begin{align}
-S^{(2)} & =\kappa_{0}\mathbb{I}+Q_{\text{TL}}\,C^{(xx)}-\frac{2}{N}Q_{\text{TL}}^{2}\,C^{(xx)}\bar{\tilde{f}}\bar{\tilde{f}}^{\T}C^{(xx)}\\
 & =\kappa_{0}\mathbb{I}+Q_{\text{TL}}\,C^{(xx)}+\mathcal{O}(1/N),
\end{align}
since $(Q\,C^{(xx)}+\kappa)\,\bar{\tilde{f}}\propto y=\mathcal{O}(1)$
and thus also $C^{(xx)}\bar{\tilde{f}}=\mathcal{O}(1)$. Here, we
drop the dependence of $Q_{\text{TL}}$ on $\bar{\tilde{f}}$ for
brevity. The fluctuation correction is given by
\begin{align}
 & \frac{1}{2}\sum_{\alpha\beta}(-\mathcal{S}^{(2)})_{\beta\alpha}^{-1}\frac{\partial^{3}(-\mathcal{S}(\bar{\tilde{f}},j_{\ast}))}{\partial\bar{\tilde{f}}_{\alpha}\partial\bar{\tilde{f}}_{\beta}\partial\bar{\tilde{f}}_{\delta}}\bigg\vert_{j^{*}=0}\\
 & =-\frac{1}{N}Q_{\text{TL}}^{2}\,\sum_{\alpha\beta}(-\mathcal{S}^{(2)})_{\beta\alpha}^{-1}\,\Big(C_{\alpha\beta}^{(xx)}\left[C^{(xx)}\bar{\tilde{f}}\right]_{\gamma}+C_{\alpha\gamma}^{(xx)}\left[C^{(xx)}\bar{\tilde{f}}\right]_{\beta}+C_{\beta\gamma}^{(xx)}\left[C^{(xx)}\bar{\tilde{f}}\right]_{\alpha}\Big)\\
 & \phantom{=}+\frac{4}{N^{2}}Q_{\text{TL}}^{3}\,\sum_{\alpha\beta}(-\mathcal{S}^{(2)})_{\beta\alpha}^{-1}\,\left[C^{(xx)}\bar{\tilde{f}}\right]_{\alpha}\left[C^{(xx)}\bar{\tilde{f}}\right]_{\beta}\left[C^{(xx)}\bar{\tilde{f}}\right]_{\gamma}
\end{align}
Looking at the individual terms, we have for the first term in the
second line
\begin{equation}
\frac{1}{N}Q_{\text{TL}}^{2}\,\text{Tr}\left[(\kappa_{0}\mathbb{I}+Q_{\text{TL}}C^{(xx)}+\mathcal{O}(1/N))^{-1}\,C^{(xx)}\right]C_{\gamma\delta}^{(xx)}\bar{\tilde{f}}_{\delta}=\mathcal{O}(P/N),
\end{equation}
where the factor $P$ results from the appearing trace. Assuming the
regularization noise $\kappa_{0}$ to be small compared to the kernel
$C^{(xx)}$, we see for the other terms that they scale as
\begin{align}
\frac{1}{N}Q_{\text{TL}}^{2}\,\sum_{\alpha\beta}(\kappa_{0}\mathbb{I}+Q_{\text{TL}}C^{(xx)}+\mathcal{O}(1/N))_{\beta\alpha}^{-1}\left(C_{\alpha\gamma}^{(xx)}C_{\beta\delta}^{(xx)}\bar{\tilde{f}}_{\delta}+C_{\beta\gamma}^{(xx)}C_{\alpha\delta}^{(xx)}\bar{\tilde{f}}_{\delta}\right) & \approx\frac{4}{N}Q_{\text{TL}}C_{\gamma\delta}^{(xx)}\bar{\tilde{f}}_{\delta}=\mathcal{O}(1/N),\\
\frac{4}{N^{2}}Q_{\text{TL}}^{3}\,\sum_{\alpha\beta}(\kappa_{0}\mathbb{I}+Q_{\text{TL}}C^{(xx)}+\mathcal{O}(1/N))_{\beta\alpha}^{-1}\,\left[C^{(xx)}\bar{\tilde{f}}\right]_{\alpha}\left[C^{(xx)}\bar{\tilde{f}}\right]_{\beta}\left[C^{(xx)}\bar{\tilde{f}}\right]_{\gamma} & =\mathcal{O}(P/N^{2}).
\end{align}
In the proportional limit $P\propto N\rightarrow\infty$, only the
first term does not vanish and the self-consistency equation for $\bar{\tilde{f}}$
becomes
\begin{equation}
iy+\kappa_{0}\bar{\tilde{f}}+Q_{\text{TL}}\big(\bar{\tilde{f}}\big)\,C^{(xx)}\bar{\tilde{f}}-\frac{1}{N}Q_{\text{TL}}^{2}\big(\bar{\tilde{f}}\big)\,\text{Tr}\left[(\kappa_{0}\mathbb{I}+Q_{\text{TL}}\big(\bar{\tilde{f}}\big)\,C^{(xx)})^{-1}C^{(xx)}\right]C^{(xx)}\bar{\tilde{f}}\overset{!}{=}0,
\end{equation}
yielding
\begin{equation}
i\bar{\tilde{f}}=\Big(\kappa_{0}\mathbb{I}+Q_{\text{TL}}\big(\bar{\tilde{f}}\big)\,\Big(1-\frac{1}{N}Q_{\text{TL}}\big(\bar{\tilde{f}}\big)\,\text{Tr}\big[(\kappa_{0}\mathbb{I}+Q_{\text{TL}}\big(\bar{\tilde{f}}\big)\,C^{(xx)})^{-1}C^{(xx)}\big]\Big)C^{(xx)}\Big)^{-1}y.\label{eq:self_cons_eq_oneloop_approx}
\end{equation}
The rescaling factor is thus given by 
\begin{equation}
Q_{\text{1-Loop}}\coloneqq Q_{\text{TL}}-\frac{1}{N}Q_{\text{TL}}^{2}\text{Tr}\big[(\kappa_{0}\mathbb{I}+Q_{\text{TL}}C^{(xx)})^{-1}C^{(xx)}\big],\label{eq:rescaling_factor_oneloop}
\end{equation}
where $Q_{\text{TL}}=Q_{\text{TL}}\big(\bar{\tilde{f}}\big)$ depends
on the self-consistent solution in \prettyref{eq:self_cons_eq_oneloop_approx}.
The tree-level solution is the leading term here and receives a correction
due to the output fluctuations. We cannot directly compare the expression
for this rescaling factor to the one in \cite{Li21_031059}, since
the latter is given by the self-consistency equation \prettyref{eq:S_Q_z-1}
and the former by the self-consistency equation \prettyref{eq:self_cons_eq_oneloop_approx}
for the training discrepancies inserted into \prettyref{eq:rescaling_factor_oneloop}.
Nevertheless, \prettyref{fig:scatter_standard_scaling_nngp} in the
main text shows that these two expressions yield the same value and
thus the same predictions for the mean discrepancies numerically.

\section{Other scaling regimes\label{app:rel_vanMeegen_somp}}

We here expose the relation of our theory to the work by \citet{vanMeegen_24_16689}
where yet another scaling regime is considered. We start from the
effective action in tree-level approximation \eqref{eq:Appendix_Tree_level_VGF}
and the form for $W$ given by \eqref{eq:cum_gen_scaling_form} 
\begin{align}
\Gamma_{\text{TL}}(\bar{\tilde{f}},j_{\ast}\vert y) & =iy_{\alpha}\bar{\tilde{f}}_{\alpha}+\frac{\kappa_{0}}{2}N^{-1}\bar{\tilde{f}}_{\alpha}\bar{\tilde{f}}_{\alpha}-W(i\bar{\tilde{f}},j_{\ast})\,\label{eq:Gamma_TL_super_mean_field-1}\\
W(\tilde{f}_{\mathcal{D}},\tilde{f}_{\ast}) & =N\,\ln\bigg\langle\int dw\,\exp\bigg(\sum_{a=1}^{P+1}\tilde{f_{a}}\frac{w}{N}\,\phi(h_{a})-P\,\frac{w^{2}}{2g_{w}}\bigg)\bigg\rangle_{h_{a}\sim\N(0,C^{(xx)})}+\const.,\nonumber 
\end{align}
where, as in the original work, we scale the variance of the readout
weights by an additional factor $P$ (see ref. \cite{vanMeegen_24_16689},
their Section ``1. Weight Posterior'' second last paragraph). In
the proportional limit $P\propto N$ it entails $w\propto N^{-\frac{3}{2}}$,
which scales down the readout weights even more strongly than mean-field
scaling. This scaling allows taking the integral over $w$ in saddle
point approximation because the exponent $\sum_{a=1}^{P+1}\tilde{f_{a}}w\,\phi(h_{a})-P\,w^{2}/2g_{w}$
scales with $P$. With \eqref{eq:appendix_Large_deviation_principle}
one therefore has with $l=(\bar{\tilde{f}},j_{\ast})/N$

\begin{align}
-\ln\,p(y|C^{(xx)})/N & =\Gamma_{\text{TL}}(N\,l\,\vert y)/N\label{eq:approx_w_star-1}\\
 & =iy^{\T}l+\frac{\kappa_{0}}{2}\,l^{\T}l\nonumber \\
 & -\ln\,\int dw\,\Big\langle\exp\Big(-P\,\frac{w^{2}}{2g_{w}}+w\,\big[il^{\T}\phi(h)\big]\Big)\Big\rangle_{h\sim\N(0,C^{(xx)})}+\const.\,\nonumber \\
 & \stackrel{P\propto N}{\simeq}iy^{\T}l+\frac{\kappa_{0}}{2}\,l^{\T}l\nonumber \\
 & -\mathrm{extr}_{w}\Big\{-P\,\frac{w^{2}}{2g_{w}}+\ln\Big\langle\exp\Big(w\,\big[il^{\T}\phi\big]\Big)\Big\rangle_{h\sim\N(0,C^{(xx)})}\Big\}+\const.\nonumber 
\end{align}
The extremum condition may have multiple degenerate values $w_{\gamma}$
that appear with the relative log probability given by the entropy
\begin{align}
\ln\,p_{\gamma} & =-P\,\frac{w_{\gamma}^{2}}{2g_{w}}+\ln\,\Big\langle\exp\Big(w_{\gamma}\,\big[il^{\ast\T}\phi\big]\Big)\Big\rangle_{h\sim\N(0,C^{(xx)})}+\const.,
\end{align}
which corresponds to Eq. (B10) in \cite{vanMeegen_24_16689} and the
constant is determined such that $\sum_{\gamma}p_{\gamma}=1$. The
distribution of $w$ hence approximates the posterior $p(w)\simeq\sum_{\gamma}p_{\gamma}\,\delta(w-w_{\gamma}^{\ast})$,
which is used in \eqref{eq:approx_w_star-1} to obtain the equation
of state \eqref{eq:eq_state_tilde_f} for $\bar{\tilde{f}}$ by taking
a partial derivative of by $l_{\alpha}$, which yields
\begin{align}
y_{\alpha} & =\sum_{\gamma}\,p_{\gamma}\,w_{\gamma}^{\ast}\,\big[\phi_{\alpha}\big]_{l^{\ast},w^{\ast}}+\kappa_{0}il^{\ast}\,,
\end{align}
which corresponds to Eq. (B17) in \cite{vanMeegen_24_16689}, where
the expectation value $\big[\ldots\big]_{l^{\ast},w^{\ast}}$ is given
by
\begin{align}
[\ldots]_{l^{\ast},w_{\gamma}^{\ast}} & :=\frac{\int d^{P}h\,\ldots\,\exp\Big(w_{\gamma}^{\ast}\,\big[il^{\ast\T}\phi(h)\big]-\frac{1}{2}h^{\T}C^{(xx)}h\Big)}{\int d^{P}h\,\exp\Big(w_{\gamma}^{\ast}\,\big[il^{\ast\T}\phi(h)\big]-\frac{1}{2}h^{\T}C^{(xx)}h\Big)}\,.
\end{align}
The extremum condition appearing in \eqref{eq:approx_w_star-1} leads
to stationary values of the weights:
\begin{align}
w_{\gamma}^{\ast} & =\frac{g_{w}}{P}\,il^{\ast\T}\,\big[\phi(h)\big]_{l^{\ast},w_{\gamma}^{\ast}}\,,
\end{align}
which corresponds to eq. B16 of \cite{vanMeegen_24_16689}.

In summary, the difference to our work is that in the scaling considered
in the work by \citet{vanMeegen_24_16689}, $w\propto1/(N\sqrt{P})$,
the readout weights $w$ tend to concentrate to non-fluctuating values
$w_{\gamma}^{\ast}$, while in mean-field scaling $w\propto1/N$,
which we treat in the main part of our work, only the network outputs
concentrate. Compared to the standard scaling $w\propto1/N$ both
scalings only require a tree-level approximation to describe the mean
predictor whereas the fluctuations in the standard scaling require
the inclusion of one-loop corrections.

\section{Details of experiments\label{app:details_experiments}}

\subsection{Self-consistency equations for numerics}

In \prettyref{app:theory}, we derive train and test statistics in
a framework involving imaginary variables $\bar{\tilde{f}}$. To solve
the resulting self-consistency equations, we need to account for their
imaginary nature and substitute in all of the results above $\bar{\tilde{f}}\rightarrow i\bar{\tilde{f}}$,
changing various signs in the process. The final expressions read
as follows: In tree-level approximation, we have
\begin{align}
\bar{\tilde{f}} & =N^{\gamma-1}\Big(\kappa_{0}\mathbb{I}+Q_{\text{TL}}\big(\bar{\tilde{f}}_{\alpha}\big)C^{(xx)}\Big)^{-1}y,\\
Q_{\text{TL}}\big(\bar{\tilde{f}}\big) & =\frac{g_{w}}{1-\frac{g_{w}}{N^{\gamma}}\bar{\tilde{f}}^{\T}C^{(xx)}\bar{\tilde{f}}},\\
\langle\Delta\rangle_{\text{TL}} & =\kappa_{0}N^{1-\gamma}\bar{\tilde{f}}=\kappa_{0}\Big(\kappa_{0}\mathbb{I}+Q_{\text{TL}}\big(\bar{\tilde{f}}\big)\,C^{(xx)}\Big)^{-1}y,\\
\langle f_{\ast}\rangle_{\text{TL}} & =Q_{\text{TL}}\big(\bar{\tilde{f}}\big)\,C_{*\alpha}^{(xx)}\Big(\kappa_{0}\mathbb{I}+Q_{\text{TL}}\big(\bar{\tilde{f}}\big)\,C^{(xx)}\Big)_{\alpha\beta}^{-1}y_{\beta}.
\end{align}
In one-loop approximation, we have for the train discrepancies
\begin{align}
\bar{\tilde{f}}_{\delta} & =\big[A\big(\bar{\tilde{f}}\big)\big]_{\delta\epsilon}^{-1}\bigg[y_{\epsilon}+\frac{1}{2}\sum_{\alpha\beta}(-\mathcal{S}^{(2)})_{\beta\alpha}^{-1}\frac{\partial^{3}(-\mathcal{S})}{\partial\bar{\tilde{f}}_{\alpha}\partial\bar{\tilde{f}}_{\beta}\partial\bar{\tilde{f}}_{\epsilon}}\bigg],\\
A\big(\bar{\tilde{f}}\big) & =\kappa_{0}\mathbb{I}+Q_{\text{TL}}\big(\bar{\tilde{f}}\big)\,C^{(xx)},\\
\frac{\partial^{2}(-\mathcal{S})}{\partial\bar{\tilde{f}}_{\alpha}\partial\bar{\tilde{f}}_{\beta}}\bigg\vert_{j^{*}=0} & =A\big(\bar{\tilde{f}}\big)+\frac{2}{N}Q_{\text{TL}}^{2}\big(\bar{\tilde{f}}\big)\,C^{(xx)}\bar{\tilde{f}}\bar{\tilde{f}}^{\T}C^{(xx)},\\
\frac{\partial^{3}(\mathcal{-S})}{\partial\bar{\tilde{f}}_{\alpha}\partial\bar{\tilde{f}}_{\beta}\partial\bar{\tilde{f}}_{\delta}}\bigg\vert_{j^{*}=0} & =\frac{2}{N}Q_{\text{TL}}^{2}\big(\bar{\tilde{f}}\big)\,\Big(C_{\alpha\beta}^{(xx)}\left[C^{(xx)}\bar{\tilde{f}}\right]_{\delta}+C_{\alpha\delta}^{(xx)}\left[C^{(xx)}\bar{\tilde{f}}\right]_{\beta}+C_{\beta\delta}^{(xx)}\left[C^{(xx)}\bar{\tilde{f}}\right]_{\alpha}\Big)\\
 & \phantom{=}+\frac{8}{N^{2}}Q_{\text{TL}}^{3}\big(\bar{\tilde{f}}\big)\,\left[C^{(xx)}\bar{\tilde{f}}\right]_{\alpha}\left[C^{(xx)}\bar{\tilde{f}}\right]_{\beta}\left[C^{(xx)}\bar{\tilde{f}}\right]_{\delta},\nonumber \\
\langle\Delta_{\alpha}\rangle_{\text{1-Loop}} & =\kappa_{0}N^{1-\gamma}\bar{\tilde{f}}_{\alpha},
\end{align}
and for the test predictors
\begin{align}
\langle f^{*}\rangle_{\mathrm{\text{1-Loop}}} & =Q_{\text{TL}}\big(\bar{\tilde{f}}\big)\,C_{*\alpha}^{(xx)}\bar{\tilde{f}}_{\alpha}-\frac{1}{2}\sum_{\alpha\beta}(-\mathcal{S}^{(2)})_{\beta\alpha}^{-1}\frac{\partial^{3}(-\mathcal{S})}{\partial\bar{\tilde{f}}_{\alpha}\partial\bar{\tilde{f}}_{\beta}\partial j_{\ast}}\bigg\vert_{j^{*}=0}.\\
\frac{\partial^{3}(\mathcal{-S})}{\partial\bar{\tilde{f}}_{\alpha}\partial\bar{\tilde{f}}_{\beta}\partial j_{\ast}}\bigg\vert_{j^{*}=0} & =\frac{2}{N}Q_{\text{TL}}^{2}\big(\bar{\tilde{f}}\big)\,\Big(C_{\alpha\beta}^{(xx)}\left[C^{(xx)}\bar{\tilde{f}}\right]_{\ast}+C_{\alpha\ast}^{(xx)}\left[C^{(xx)}\bar{\tilde{f}}\right]_{\beta}+C_{\beta\ast}^{(xx)}\left[C^{(xx)}\bar{\tilde{f}}\right]_{\alpha}\Big)\\
 & \phantom{=}+\frac{8}{N^{2}}Q_{\text{TL}}^{3}\big(\bar{\tilde{f}}\big)\,\left[C^{(xx)}\bar{\tilde{f}}\right]_{\alpha}\left[C^{(xx)}\bar{\tilde{f}}\right]_{\beta}\left[C^{(xx)}\bar{\tilde{f}}\right]_{\ast},\nonumber 
\end{align}
Finally, in the proportional limit $P\propto N\rightarrow\infty$
this reduces to
\begin{align}
\bar{\tilde{f}} & =\Big(\kappa_{0}\mathbb{I}+Q_{\text{1-Loop}}\big(\bar{\tilde{f}}\big)\,C^{(xx)}\Big)^{-1}y,\\
Q_{\text{1-Loop}}\big(\bar{\tilde{f}}\big) & =Q_{\text{TL}}\big(\bar{\tilde{f}}\big)-\frac{1}{N}Q_{\text{TL}}^{2}\big(\bar{\tilde{f}}\big)\,\text{Tr}\big[(\kappa_{0}\mathbb{I}+Q_{\text{TL}}\big(\bar{\tilde{f}}\big)\,C^{(xx)})^{-1}C^{(xx)}\big],\\
\langle\Delta\rangle & =\kappa_{0}N^{1-\gamma}\Big(\kappa_{0}\mathbb{I}+Q_{\text{1-Loop}}\big(\bar{\tilde{f}}\big)\,C^{(xx)}\Big)^{-1}y,\\
\langle f^{*}\rangle_{\mathrm{\text{1-Loop, rescaling}}} & =\big[Q_{\text{1-Loop}}\big(\bar{\tilde{f}}\big)\,C_{\ast\alpha}^{(xx)}\big]\,\big(\kappa_{0}\I+Q_{\text{1-Loop}}\big(\bar{\tilde{f}}\big)\,C^{(xx)}\big)_{\alpha\beta}^{-1}\,y_{\beta}.
\end{align}

\subsection{Numerical stability and computational complexity}

To improve the numerical stability when solving these self-consistency
equations, we use the following scheme: In standard scaling, we use
the train predictors given by the NNGP as starting values for solving
the tree-level equations. Then, we use the tree-level solution as
the initial value for the one-loop equations since the latter contain
additional fluctuation corrections to the tree-level result. Further,
we anneal solutions from the standard to the mean-field regime since
the solutions are more unstable in the mean-field regime (see pseudo
code in \ref{alg:scale_annealing}).\begin{algorithm}[tb]
\caption{Annealing of solutions across scaling regimes}
\label{alg:scale_annealing}
\begin{algorithmic}
   \STATE {\bfseries Input:} data $X$, labels $Y$, scales $\{\chi_i\}_i$
   \STATE Compute NNGP train predictors $f_{\alpha}^{\text{NNGP}}$ from data $X$ and labels $Y$.
   \STATE Set initial value to NNGP predictor $f_{\alpha}^{\text{NNGP}}$.
   \FOR{$\chi$ {\bfseries in} $\{\chi_i\}_i$}
   \STATE Set $g_w\mapsto g_w/\chi$.
   \STATE Solve self-consistency solution for tree-level approximation $\bar{\tilde{f}}_{\alpha}^{\text{TL}}$ with initial value $\bar{\tilde{f}}_{\alpha}^{\text{NNGP}}$.
   \STATE Solve self-consistency solution for one-loop approximation $\bar{\tilde{f}}_{\alpha}^{\text{1-Loop}}$ with initial value $\bar{\tilde{f}}_{\alpha}^{\text{TL}}$.   \ENDFOR
\end{algorithmic}
\end{algorithm}

The computational complexity for solving these equations is $\mathcal{O}(P^{3})$
with a different pre-factor for tree-level and one-loop. Note that
neither the input dimension $D$ nor the network width $N$ but only
the number of training samples $P$ affect the computational complexity.

\subsection{Network tasks and training\label{app:tasks_training}}

\paragraph*{Ising task}

We use a linearly separable Ising task: Each pattern $x_{\alpha}$
in the Ising task is $D$-dimensional and $x_{\alpha i}\in\{\pm1\}$.
If the pattern belongs to class $-1$, each $x_{\alpha i}$ realizes
$x_{\alpha i}=+1$ with a probability of $p_{1}=0.5-\Delta p$ and
the value $x_{\alpha i}=-1$ with $p_{2}=0.5+\Delta p$. The value
for each pattern element $x_{\alpha i}$ is drawn independently. If
the pattern belongs to class $+1$, the probabilities for $x_{\alpha i}=1$
and $x_{\alpha i}=-1$ are inverted. The task complexity decreases
with larger $\Delta p$. We use $\Delta p=0.1$ throughout, corresponding
to an oracle accuracy on the classification task of $P_{\text{oracle}}=99,78\%$.

\paragraph*{Teacher-student task}

In this setting, the target is given by a $y_{\alpha}=w_{*}\cdot x_{\alpha}$,
where $x_{\alpha}\in\mathbb{R}^{D}$ is standard normally distributed
$x_{\alpha i}\sim\mathcal{N}(0,\mathbb{I})$. The teacher direction
$w_{*}\in\mathbb{R}^{D}$ is chosen to be $\hat{e}_{1}$ in the standard
basis.

\paragraph*{Network training}

We train networks using Langevin stochastic gradient descent (LSGD)
as detailed in \cite{Naveh21_064301} so that the trained networks
are effectively sampled from the posterior distribution \prettyref{eq:dist_output}.
Here evolving network parameters $\Theta$ such as weights $V,w$
with the stochastic differential equation
\begin{align}
\partial_{t}\Theta(t) & =-\rho\Theta(t)-\nabla_{\Theta}\mathcal{L}(\Theta(t);y)+\sqrt{2T}\zeta(t),\label{eq:Appendix_SingleParameter_LSGD}\\
\big\langle\zeta_{i}(t)\zeta_{j}(s)\big\rangle & =\delta_{ij}\delta(t-s),\nonumber 
\end{align}
with the squared error loss $\mathcal{L}(\Theta;y)=\sum_{\alpha=1}^{P}(f_{\alpha}(\Theta)-y_{\alpha})^{2}$,
$\zeta$ a unit variance Gaussian white noise, and $f_{\alpha}(\Theta)$
denoting the network output for sample $\alpha=1,\ldots,P$, leads
to sampling from the equilibrium distribution for $\Theta$ for large
times $t$ which reads
\begin{equation}
\lim_{t\rightarrow\infty}p\left(\Theta(t)\right)\sim\exp\left(-\frac{\rho}{2T}\|\Theta\|^{2}-\frac{1}{T}\mathcal{L}(\Theta;y)\right).\label{eq:Appendix_equilibrated_dist}
\end{equation}
Using the Fokker-Planck equation \cite{Risken96} one can derive this
density for $\Theta$. Further, this implies a distribution on the
network output
\begin{align}
p(Y|X)\propto & \int d\Theta\,\exp\big(-\frac{\rho}{2T}\,\|\Theta\|^{2}-\frac{1}{T}\,\|f-y\|^{2}\big)\\
\propto & \Big\langle\exp\big(-\frac{1}{T}\,\|f-y\|^{2}\big)\Big\rangle_{\Theta_{k}\stackrel{\text{i.i.d.}}{\sim}\N(0,T/\rho)}\nonumber \\
\propto & \N(y|f,T/2)\,\langle\delta\big[f-f(\Theta)\big]\rangle_{\Theta_{k}\stackrel{\text{i.i.d.}}{\sim}\N(0,T/\rho)},\nonumber 
\end{align}
In fact, $p(f|X)\equiv\langle\delta\big[f-f(\Theta)\big]\rangle_{\Theta_{k}\stackrel{\text{i.i.d.}}{\sim}\N(0,T/\rho)}$,
leads to the posterior in \prettyref{eq:cum_gen_posterior} if one
identifies $\kappa_{0}=T/2$ with the regularization noise and $T/\ensuremath{\rho}=g/N$
with the variance of the parameter $\Theta_{k}$. Implementing the
sampling in practice this corresponds to requiring different weight
decay $\rho$ for each parameter, as weight variances can differ in
the input and output layer.

The time discrete version of \prettyref{eq:Appendix_SingleParameter_LSGD}
is implemented in our PyTorch code as 
\begin{align}
\Theta_{t} & =\Theta_{t-1}-\eta\left(\rho\Theta_{t-1}+\nabla_{\Theta}\mathcal{L}(\Theta_{t-1};y)\right)+\sqrt{2T\eta}\,\zeta_{t},\\
\langle\zeta_{t}\zeta_{s}\rangle & =\delta_{ts},\nonumber 
\end{align}
with standard normal $\zeta_{t}$ and finite time step $\eta$, which
can also be interpreted as a learning rate. To accurately reflect
the time evolution according to \prettyref{eq:Appendix_SingleParameter_LSGD}
the learning rate $\eta$ needs to be small enough.

Hence the LSGD we implement corresponds to full-batch gradient descent
with the addition of i.i.d. distributed standard normal noise and
weight decay regularization \cite{Krogh91}. The value for $\kappa_{0}$
corresponds to a tradeoff in the optimization between the weight priors
and the likelihood in terms of the loss $\mathcal{L}$. Choosing large
$\kappa_{0}$ corresponds to large $T=2\kappa_{0}$ and hence a large
noise in the LSGD and therefore putting more emphasis on the Gaussian
parameter priors. Small regularization values $\kappa_{0}$ favor
the training data in terms of the loss in the exponent of \eqref{eq:Appendix_equilibrated_dist}.

To faithfully compare the numerical results with our theoretical results,
the LSGD needs to sample from the equilibrium distribution. For this
it needs to be ensured that the distribution is equilibrated by evolving
the networks for $10.000$ steps . We ensure uncorrelated network
samples by initializing different networks with different random seeds.

For the Ising task, we average over $N_{\text{networks}}=100$ with
different initial weights to obtain the training and test predictors.
For the teacher-student task, we average over $N_{\text{networks}}=5.000$
with different initial weights to obtain the covariance of the network
output projected onto different directions.

\section{Additional figures\label{app:add_figures}}

\subsection{MNIST}

Since the presented approach does not make any assumption on the data,
it is applicable to arbitrary data sets. We here show results for
binary classification on MNIST across scaling regimes.

\begin{figure}[H]
\begin{centering}
\includegraphics[width=0.6\textwidth]{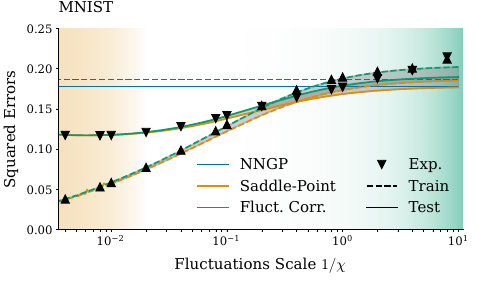}\vspace{0.2in}
\par\end{centering}
\centering{}\caption{Binary classification on MNIST: training (solid line) and test errors
(dashed line) across scaling regimes for different approaches. While
standard scaling (green shaded area) requires a one-loop approximation
with fluctuation corrections (Fluct. Corr.), a saddle-point or tree-level
approximation (Saddle-Point) is sufficient in mean-field scaling (orange
shaded area). Parameters: $P_{\text{train}}=80$, $N=100$, $D=784$,
$\kappa_{0}=1$, $P_{\text{test}}=10^{3}$, $g_{v}=g_{w}=2$.}
\label{fig:mnist_transition}\vspace{0.2in}
\end{figure}
\newpage

\begin{figure}[H]
\vskip 0.2in
\begin{centering}
\includegraphics[width=1\textwidth]{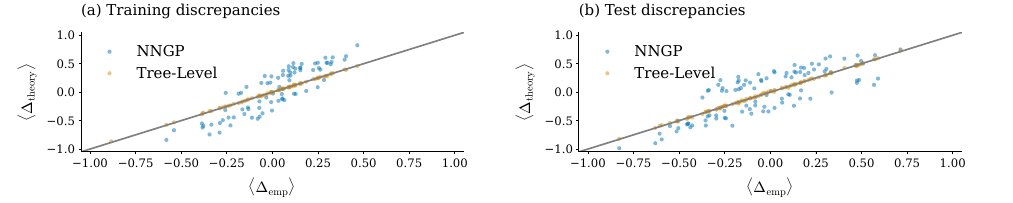}\caption{(a) Training discrepancies $\langle\Delta\rangle=y-\langle f_{\mathcal{D}}\rangle$
and (b) test discrepancies $\langle\Delta_{\ast}\rangle=y_{\ast}-\langle f_{\ast}\rangle$
for binary classification on MNIST in mean-field scaling. We show
theoretical values for both NNGP and tree-level against empirical
results, where the gray line marks the identity. In contrast to the
NNGP, the tree-level approximation accurately matches the empirical
values. Parameters: $\gamma=2$, $P_{\text{train}}=80$, $N=100$,
$D=784$, $\kappa_{0}=1$, $P_{\text{test}}=10^{3}$, $g_{v}=g_{w}=2$.}
\label{fig:mnist_scatter_mean_field}
\par\end{centering}
\vskip -0.2in
\end{figure}

\begin{figure}[H]
\vskip 0.2in
\begin{centering}
\includegraphics[width=1\textwidth]{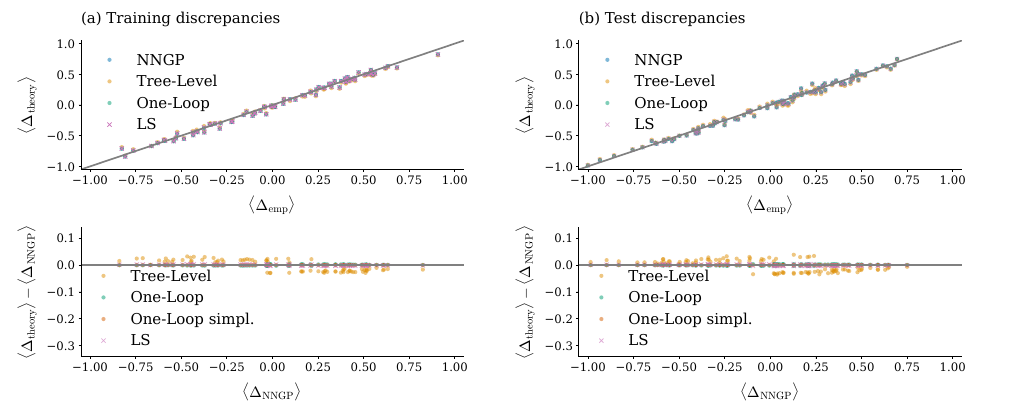}
\par\end{centering}
\begin{centering}
\caption{(a) Training discrepancies $\langle\Delta\rangle=y-\langle f_{\mathcal{D}}\rangle$
and (b) test discrepancies $\langle\Delta_{\ast}\rangle=y_{\ast}-\langle f_{\ast}\rangle$
for binary classification on MNIST in standard scaling. Upper row:
theoretical values for different theories against empirical results;
gray line marks the identity. Lower row: difference of theoretical
values to the NNGP as a baseline against NNGP predictions, indicating
small-scale differences between the different approaches. Results
of the kernel approach by \citet{Li21_031059} shown as reference
(LS). Parameters: $\gamma=1$, $P_{\text{train}}=80$, $N=100$, $D=784$,
$\kappa_{0}=1$, $P_{\text{test}}=10^{3}$, $g_{v}=g_{w}=2$.}
\label{fig:mnist_scatter_standard_scaling_nngp}
\par\end{centering}
\vskip -0.2in
\end{figure}
\newpage

\subsection{Ising task in mean-field scaling}

\begin{figure}[H]
\begin{centering}
\includegraphics[width=1\textwidth]{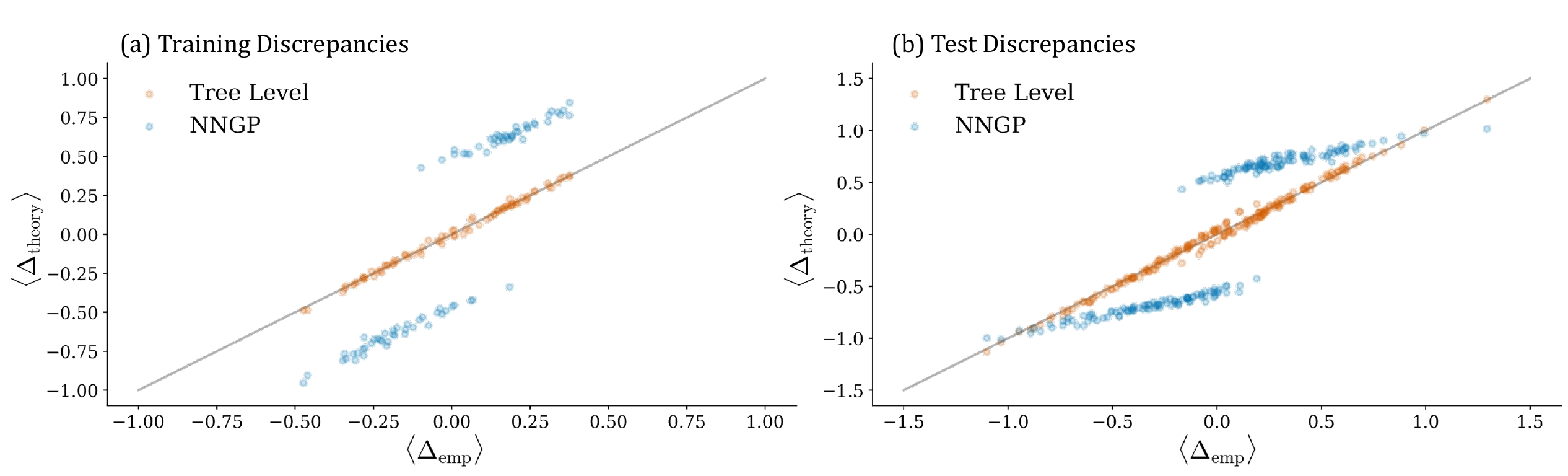}\vspace{0.2in}
\par\end{centering}
\centering{}\caption{(a) Training discrepancies $\langle\Delta\rangle=y-\langle f_{\mathcal{D}}\rangle$
and (b) test discrepancies $\langle\Delta_{\ast}\rangle=y_{\ast}-\langle f_{\ast}\rangle$
on an Ising task in mean-field scaling. We show theoretical values
for both NNGP and tree-level against empirical results, where the
gray line marks the identity. In contrast to the NNGP, the tree-level
approximation accurately matches the empirical values. We here use
a non-linear activation function $\phi=\text{erf}$. Parameters: $\gamma=2$,
$P_{\text{train}}=80$, $N=100$, $D=200$, $\kappa_{0}=1$, $P_{\text{test}}=10^{3}$,
$g_{v}=g_{w}=0.5$, $\Delta p=0.1$.\label{fig:nonlinear_scatter}}
\vspace{0.2in}
\end{figure}

\begin{figure}[H]
\vskip 0.2in
\begin{centering}
\includegraphics[width=1\textwidth]{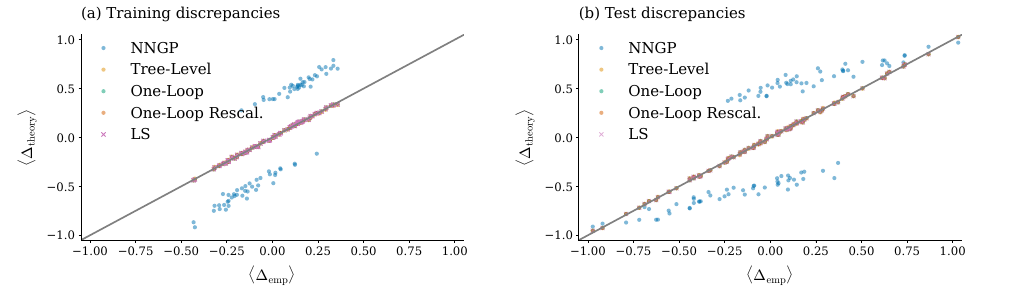}\caption{Scatter plots of (a) training discrepancies $\langle\Delta_{\alpha}\rangle=y_{\alpha}-\langle f_{\alpha}\rangle$
and (b) test discrepancies $\langle\Delta_{\ast}\rangle=y_{\ast}-\langle f_{\ast}\rangle$
on an Ising task in mean-field scaling. We show theoretical values
for NNGP and different feature learning theories against empirical
results, where the gray line marks the identity. In contrast to the
NNGP, the tree-level approximation accurately matches the empirical
values. Further, the different feature learning theories lie on top
of one another in mean-field scaling. Parameters: $\gamma=2$, $P_{\text{train}}=80$,
$N=100$, $D=200$, $\kappa_{0}=0.4$, $P_{\text{test}}=10^{3}$,
$g_{v}=0.5$, $g_{w}=0.2$, $\Delta p=0.1$.}
\label{fig:scatter_mean_field-1}
\par\end{centering}
\vskip -0.2in
\end{figure}
\newpage

\subsection{Coherent amplification of low-rank kernel structures}

\begin{figure}[H]
\begin{centering}
\includegraphics[width=1\textwidth]{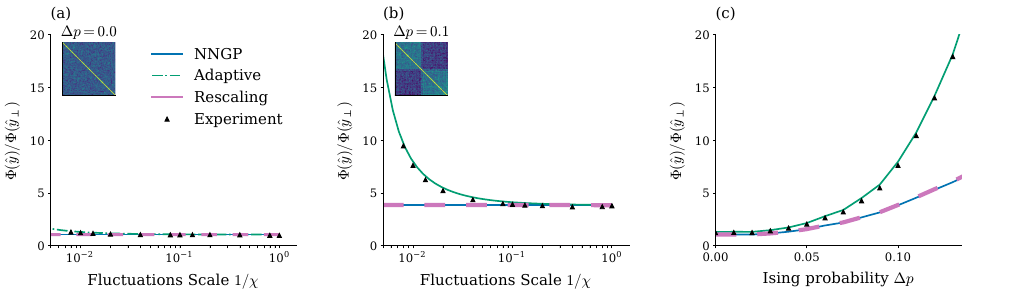}\vspace{0.2in}
\par\end{centering}
\centering{}\caption{Relative directional feature learning on the Ising task as a function
of the fluctuation scale $1/\chi$ for (a) an input kernel without
structure and (b) an input kernel with block structure (input kernels
shown as insets). Both NNGP and rescaling theory fail to capture directional
feature learning, while the multi-scale adaptive theory accurately
predicts network behavior. (c) An increase in structure in the input
kernel increases with the Ising probability $\Delta p$ and leads
to a significantly higher directional feature learning in the adaptive
theory than in both NNGP and rescaling, matching the experiments.
Parameters: $P_{\text{train}}=80$, $N=100$, $D=200$, $\kappa_{0}=2$,
$g_{v}=0.01,$ $g_{w}=0.5$.\label{fig:Relative-directional-feature}}
\vspace{0.2in}
\end{figure}

\subsection{Teacher-student task}

\begin{figure}[H]
\vskip 0.2in
\begin{centering}
\includegraphics[width=1\textwidth]{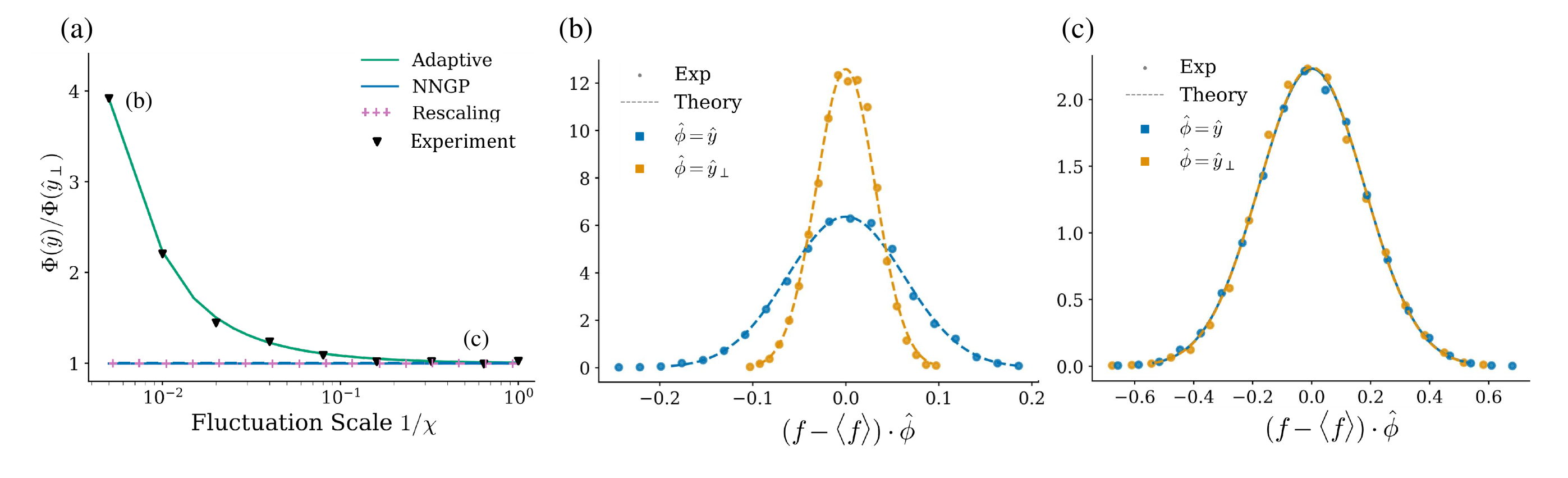}\caption{(a) Directional feature learning in a teacher-student setting as a
function of the fluctuation scale $1/\chi$. Both NNGP and rescaling
theory fail to capture directional feature learning, while the multi-scale
adaptive theory accurately predicts network behavior. Output distribution
in different directions (b) in mean field scaling ($\chi=N$) and
(c) in standard scaling ($\chi=128$) . Parameters: $P_{\text{train}}=80$,
$N=200$, $D=50$, $\kappa_{0}=2$, $g_{v}=0.01,$ $g_{w}=2$.\label{fig:teacher_student_appendix}}
\par\end{centering}
\vskip -0.2in
\end{figure}

\end{document}